\documentclass[trackchanges,twocolumn]{aastex701}

\usepackage{dashrule}

\begin{document}

\title{
    XRISM Time-resolved Fe~K$\alpha$ Spectroscopy of NGC 4395: Time-variable Inner-disk Emission 
  }

\author[orcid=0000-0002-6808-2052]{Taiki Kawamuro}
\affiliation{Department of Earth and Space Science, Graduate School of Science, The University of Osaka, 1-1 Machikaneyama, Toyonaka, Osaka 560-0043, Japan}
\affiliation{RIKEN Pioneering Research Institute (PRI), 2-1 Hirosawa, Wako, Saitama 351-0198, Japan}
\email[show]{kawamuro@ess.sci.osaka-u.ac.jp}

\author[orcid=0000-0002-9754-3081]{Satoshi Yamada} 
\affiliation{The Frontier Research Institute for Interdisciplinary Sciences, Tohoku University, Aramaki, Aoba-ku, Sendai, Miyagi 980-8578, Japan}
\affiliation{Astronomical Institute, Tohoku University, 6-3 Aramakiazaaoba, Aoba-ku, Sendai, Miyagi 980-8578, Japan}
\affiliation{Department of Astronomy, University of Geneva, Ch.d’Ecogia 16, 1290, Versoix, Switzerland}
\email[]{satoshi.yamada@astr.tohoku.ac.jp}

\author[0000-0001-6020-517X]{Hirofumi Noda}
\affiliation{Astronomical Institute, Tohoku University, 6-3 Aramakiazaaoba, Aoba-ku, Sendai, Miyagi 980-8578, Japan}
\email[]{hirofumi.noda@astr.tohoku.ac.jp}

\author[0000-0002-7272-1136]{Yoshiyuki Inoue}
\affiliation{College of Systems Engineering and Science, Shibaura Institute of Technology, 307 Fukasaku, Minuma-ku, Saitama City, Saitama 337-8570, Japan}
\affiliation{Department of Earth and Space Science, Graduate School of Science, The University of Osaka, 1-1 Machikaneyama, Toyonaka, Osaka 560-0043, Japan}
\affiliation{Interdisciplinary Theoretical \& Mathematical Science Center (iTHEMS), RIKEN, 2-1 Hirosawa, Wako, Saitama 351-0198, Japan}
\affiliation{Kavli Institute for the Physics and Mathematics of the Universe (WPI), UTIAS, The University of Tokyo, 5-1-5 Kashiwanoha, Kashiwa, Chiba 277-8583, Japan}
\email[]{yoshiyuki.inoue.sci@osaka-u.ac.jp}

\author[0000-0002-5701-0811]{Shoji Ogawa} 
\affiliation{Faculty of Science and Technology, Tokyo University of Science, 2641, Yamazaki, Noda, Chiba 278-8510, Japan}
\affiliation{Institute of Space and Astronautical Science (ISAS),
  Japan Aerospace Exploration Agency (JAXA),
  3-1-1 Yoshinodai, Chuo-ku, Sagamihara, Kanagawa 252-5210, Japan}
\email[]{sogawa@ac.jaxa.jp}

\author[orcid=0000-0003-2161-0361]{Misaki Mizumoto}
\affiliation{Science Education Research Unit, University of Teacher Education Fukuoka, Munakata, Fukuoka 811-4192, Japan}
\email[]{mizumoto-m@fukuoka-edu.ac.jp}





\begin{abstract}

We report the first XRISM observation of the low-mass AGN in the nearby dwarf galaxy NGC~4395 ($M_{\rm BH}\sim10^{4-5}\,M_\odot$), complemented by a simultaneous NuSTAR observation. We constrained the continuum by jointly fitting the XRISM/Resolve (2--12~keV) and NuSTAR (3--30~keV) spectra while excluding the Fe~K band (5.5--7.5~keV).
Relative to this baseline continuum, the time-averaged Resolve spectrum revealed an unresolved neutral Fe~K$\alpha$ core with a velocity width of $\lesssim$110~km~s$^{-1}$ and an adjacent redward wing. 
The red wing was well reproduced by an additional relativistically broadened
Fe~K component.
Furthermore, time-resolved spectroscopy with $\approx$87 ks bins showed that the diskline profile varied significantly over the $\sim$400~ks observation. 
This evolution can be interpreted in terms of changes in the inner radius of the line-emitting region, together with a possible inclination modulation with a period of $\approx$210~ks.
If interpreted as Lense--Thirring precession of a tilted inner flow, the observed period would favor the low end of the black hole mass estimates ($M_{\rm BH}\approx9\times10^3\,M_\odot$) and imply a moderate spin ($a\gtrsim0.6$). These results highlight the capability of XRISM to track relativistic disk dynamics in AGNs.

\end{abstract}




\section{Introduction} 

Low-mass active galactic nuclei (AGNs) hosting black holes (BHs) of $M_{\rm BH}\sim10^{3}$--$10^{5}\,M_\odot$ provide a unique laboratory for time-domain studies of accretion disks. Owing to their small BH masses, the characteristic disk timescales (e.g., dynamical, thermal, and viscous timescales) are much shorter than those in typical Seyfert galaxies, making it feasible to trace structural changes in the inner disk within a single observation. Moreover, if the angular momentum of the inner disk is misaligned with the BH spin axis, general-relativistic frame dragging can induce Lense--Thirring (LT) precession \citep{Lense1918PhyZ...19..156L}. Detecting LT-driven geometric modulation would provide a direct probe of the three-dimensional inner accretion geometry and could offer complementary constraints on the BH spin \citep[e.g.,][]{Fragile2024arXiv240410052F}.

Broadened, skewed Fe~K$\alpha$ line emission around 6\,keV is a direct probe of the geometry of the innermost accretion disk and has been investigated for a long time \citep[e.g.,][]{George1991MNRAS.249..352G,Tanaka1995Natur.375..659T}. 
Now, XRISM/Resolve \citep[][]{Tashiro2025PASJ...77S...1T,Ishisaki2025JATIS..11d2023I,Kelley2025JATIS..11d2026K} is beginning to play an important role in revealing the diskline emission in nearby AGNs. 
Its unprecedented spectral resolution of $\approx$4.5\,eV at 6 keV has enabled 
precise measurements of diskline profiles, 
while identifying and accounting for narrower Fe~K$\alpha$ components and absorption lines, if any \citep[e.g.,][]{Xrism2024ApJ...973L..25X,Brenneman2025ApJ...995..200B,Miller2025ApJ...994L..10M}. 
However, the time evolution of diskline features has so far been only
limitedly explored with Resolve, despite its importance for understanding
the structure and dynamics of the accretion disk
\citep[][]{Wilkins2026arXiv260409761W}.

The AGN in the nearby dwarf galaxy NGC~4395 
at $z = $ 0.001064 \citep{Haynes1998AJ....115...62H} 
is an ideal target for investigating the evolution of the inner-disk geometry 
by tracking changes in the diskline profile.  
The AGN hosts a low-mass BH ($M_{\rm BH} \sim10^{4}$--$10^{5}\,M_\odot$; e.g., \citealt{Peterson2005ApJ...632..799P,Woo2019NatAs...3..755W}) and
is X-ray bright \citep[$\sim$4--8$\times10^{-12}$~erg~cm$^{-2}$~s$^{-1}$ in the 2--10~keV band; e.g.,][]{Vaughan2005MNRAS.356..524V,Iwasawa2010A&A...514A..58I}. 
In addition, it is known to show strong variability in X-ray flux \citep[e.g.,][]{Vaughan2005MNRAS.356..524V}. 
The low BH mass gives access to short disk timescales, the X-ray brightness
enables high-quality Resolve spectroscopy, and the strong variability makes it possible to search for diskline profile evolution within a single observation.

In this paper, we present the first XRISM observation of NGC~4395, complemented by a simultaneous Nuclear Spectroscopic Telescope Array (NuSTAR; \citealt{Har13})  observation. 
We focus on the time evolution of the broadened, skewed Fe~K$\alpha$ emission and discuss possible geometric interpretations, including LT precession.
This paper is organized as follows. 
Section~\ref{sec:xrayobs} introduces our XRISM and NuSTAR observations. 
Section~\ref{sec:datana} presents 
 data reduction and analyses, including imaging (Section~\ref{sec:image}), light-curve generation (Section~\ref{sec:lc_gen}), and spectral fitting (Section~\ref{sec:spec_gen}). 
Section~\ref{sec:dis} discusses time-variable broadened Fe~K$\alpha$ emission, and Section~\ref{sec:sum} summarizes our conclusions. \\

\section{XRISM and NuSTAR  Observations}\label{sec:xrayobs} 

We performed coordinated XRISM and NuSTAR observations of NGC 4395 (PI: T. Kawamuro). 
The XRISM observation (ObsID 201048010) started on 2024 November 7 and spanned $\approx$443~ks, yielding net exposures of $\approx$237~ks and
$\approx$167~ks for Resolve and Xtend, respectively. 
The shorter exposure of Xtend is due to a data readout issue  near the end of the observation.
Resolve was used with an open filter, while Xtend was operated in full-window mode \citep{Noda2025PASJ...77S..10N}. 

Our NuSTAR observation (ObsID 91001636002) started on 2024 November 8, about 100 ks after the start of the XRISM observation. 
It lasted for $\approx$154 ks with a net exposure of $\approx$81\,ks. \\

\section{Data Analysis}\label{sec:datana} 

We analyzed the XRISM and NuSTAR data
using HEASoft v6.35.2, including NuSTAR Data Analysis Software (NuSTARDAS) v2.1.4. 
The calibration databases we adopted for Resolve, Xtend, and the NuSTAR 
focal plane detector modules (FPMs) were those released on 2025 March 15, 2024 November 15, and 2025 September 08, respectively. 

We reduced the XRISM and NuSTAR data following standard procedures. 
For XRISM/Resolve, we started from the cleaned event file created by the XRISM team (the prepipeline package ver. 005 003.20Jun2024 Build8.012 and the standard pipeline script 03.00.013.010) and then applied the standard screening and event-quality selections. 
Furthermore, we excluded events from Pixels 12 and 27 \citep{2024SPIE13093E..1PE}. 
For XRISM/Xtend, we used the distributed cleaned event file as the basis. 
For NuSTAR, we reprocessed the raw data with \texttt{nupipeline} to obtain cleaned event files for FPMA and FPMB.

\begin{figure}
  \centering
  \includegraphics[scale=0.56]{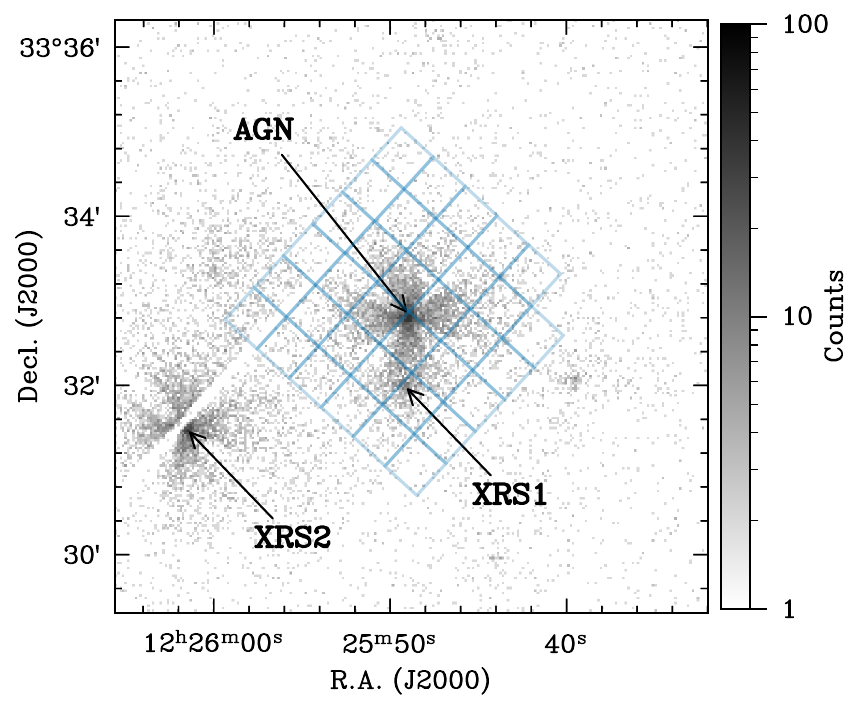}  
    \caption{
    Xtend 0.5--10\,keV count map centered on NGC~4395. 
Blue squares indicate the footprint of the Resolve pixel array. 
Black arrows mark the AGN and two nearby X-ray sources, XRS1 and XRS2.
    }  
\label{fig:xtdimg} 
\end{figure}

\subsection{Xtend Image}  \label{sec:image}

\begin{figure*}
  \centering
  \includegraphics[scale=0.5]{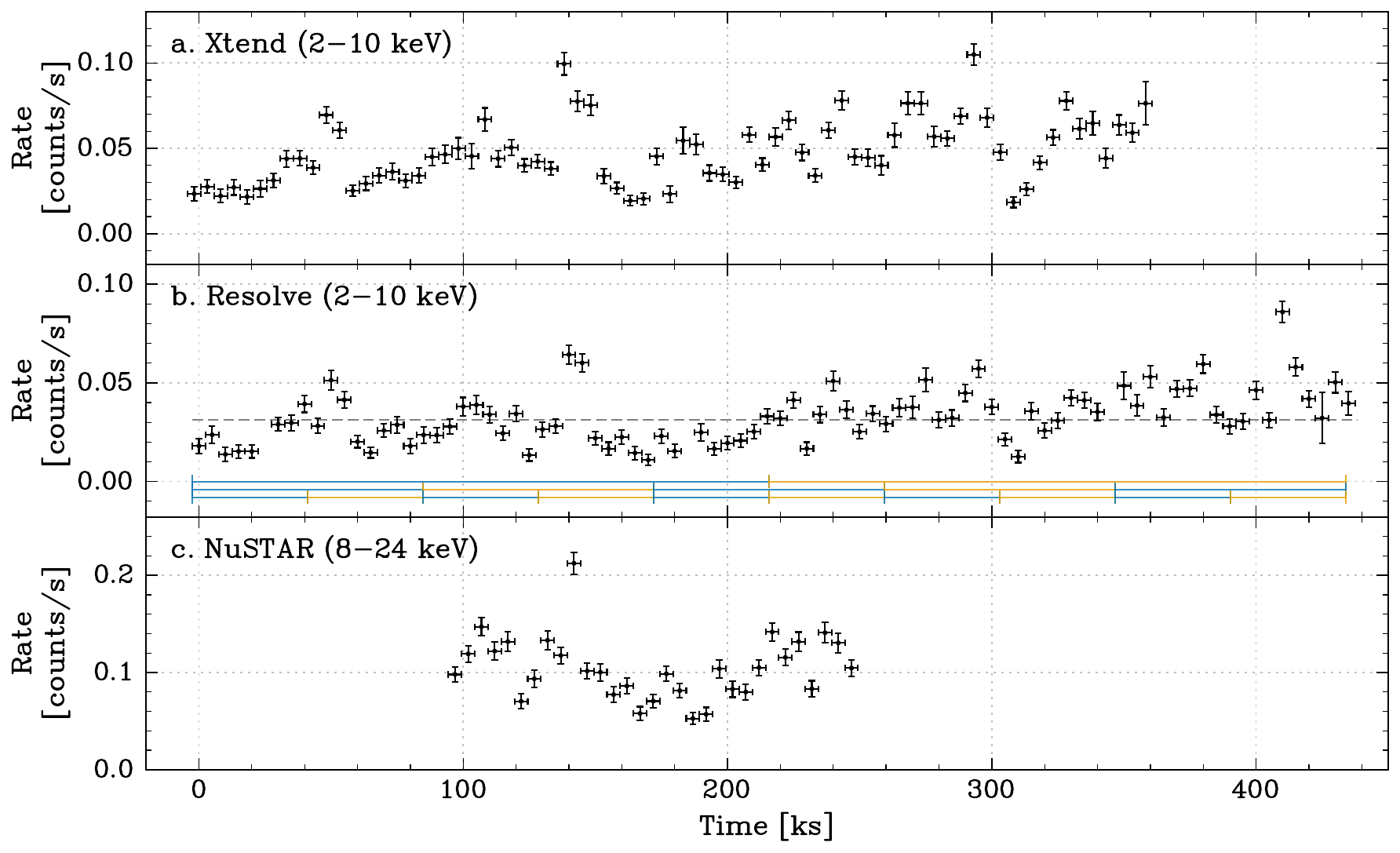}  \vspace{-0.cm}
    \caption{
    Background-subtracted Xtend (2--10\,keV), Resolve (2--10\,keV), and NuSTAR
    (8--24\,keV) light curves, binned at 5\,ks.
    In panel~b, the dashed line marks the median count rate.
    The blue and orange bars indicate the time intervals used for the
    time-resolved spectral analyses in Sections~\ref{sec:twobin} and
    \ref{sec:finebin}.
    The intervals divide the observation into 2 segments (top), 5 segments
    (middle), and 10 segments (bottom).
    }  
  \label{fig:xlc} 
\end{figure*}

Creating a 0.5--10~keV Xtend image (Figure~\ref{fig:xtdimg}), we assessed possible contamination of Resolve AGN spectra from nearby sources. 
We identified two sources at (R.A., decl.) = (12$^{\rm h}$25$^{\rm m}$49$\fs$25, +33$\degr$32$\arcmin$00$\farcs$60) and
(12$^{\rm h}$26$^{\rm m}$01$\fs$61, +33$\degr$31$\arcmin$30$\farcs$00), which we refer to as XRS1 and XRS2, respectively. 
From Xtend spectral extractions and modeling, we confirmed that the contamination from XRS1 is $<$10\% above 3~keV and that XRS2 is negligible.
We therefore decided to restrict all Resolve spectral analyses to 3--12~keV, except
for the broadband continuum fits used to determine the continuum shape
robustly (Section~\ref{sec:res_nus}).

\subsection{Light curves}  \label{sec:lc_gen} 

We produced background-subtracted light curves with 5~ks bins, as shown in Figure~\ref{fig:xlc}. 
For Xtend (2--10~keV), we used a 60\arcsec\ circular source region centered on the AGN and a 150\arcsec\ source-free background region. 
Because Resolve has no off-source region, we estimated the non-X-ray background (NXB) using \texttt{rslnxbgen}; the cosmic X-ray background is negligible compared with the NXB. 
For NuSTAR, we produced an 8--24~keV light curve with \texttt{nuproducts}, adopting 60\arcsec\ and 90\arcsec\ circular regions for the source and background, respectively.

We fitted each background-subtracted light curve with a constant
model and found that the constant model was rejected in all curves with $p$-values of $\ll$ 0.01. This confirms significant X-ray flux variability during the observation and motivates the time-resolved spectral analyses described later.

\subsection{Spectral Analysis} \label{sec:spec_gen} 

We performed spectral analysis using Resolve and NuSTAR (FPMA+FPMB). 
Resolve spectra were constructed by accumulating high-primary events for the full observation (Section~\ref{sec:time_ave_res}) and also for time intervals used in the time-resolved analyses (Sections~\ref{sec:twobin} and ~\ref{sec:finebin}). 
We generated an L-sized response matrix file (RMF)  with \texttt{rslmkrmf} and computed an auxiliary response file (ARF) using a fiducial parameter set. 
We did not subtract the NXB. Instead, we modeled it using the canonical NXB model\footnote{\url{https://heasarc.gsfc.nasa.gov/docs/xrism/analysis/nxb/nxb_spectral_models.html}} and fitted it simultaneously with the source spectrum.
NuSTAR source and background spectra, together with the corresponding
responses, were produced with \texttt{nuproducts} using the same regions as those for the light curves.

All spectra were binned using the optimal binning scheme while ensuring at least one count per bin \citep{Kaastra2016A&A...587A.151K}, and fitted in XSPEC~v12.15.1 \citep{Arn96} using the $C$ statistic \citep{Cas79}. 
To assess the validity of the best-fit models, we computed the expected $C$ values  following \citep{Kaa17}, and then compared them with the minimized $C$ values.
As a supplemental check, we examined how the relation between the minimized and expected
$C$ statistic depends on spectral binning by performing Monte Carlo simulations of a representative best-fit model (Appendix~\ref{app:binning}).

For model comparison, we used the Akaike information criterion \citep{Akaike1974ITAC...19..716A}. 
We define $\Delta{\rm AIC}$ as the AIC value of the model with an additional
component minus that of the reference model without that component.
Following the standard interpretation of AIC differences, we regard
$\Delta{\rm AIC}<-2$ as substantial evidence in favor of the model with the
additional component
\citep[e.g.,][]{Burnham_2002,Miller2025ApJ...994L..10M,Fujiwara2026arXiv260406719F}.
Uncertainties are quoted at the 90\% confidence level unless otherwise noted.

\begin{figure}
  \centering
    \includegraphics[scale=0.38]{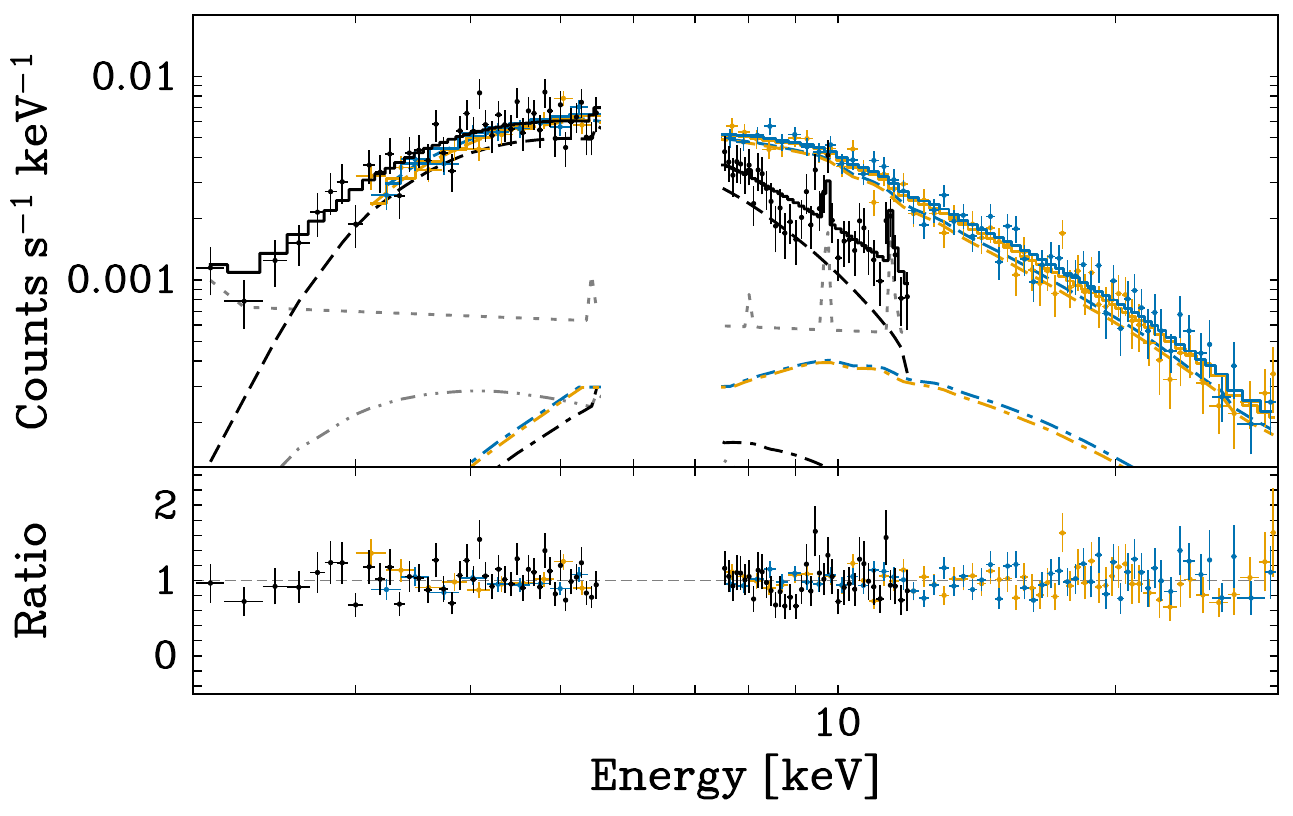}
    \caption{
    Joint continuum fit to Resolve (black), NuSTAR/FPMA (blue), and FPMB (yellow) spectra extracted during the $\sim$ 80\,ks NuSTAR observation. 
    The Fe band (5.5--7.5\,keV) is excluded in the fit. 
      Solid lines show the best-fit models; for Resolve, the dashed and dot--dashed lines indicate the power-law and NXB components, respectively. The lower panel shows data-to-model ratios.
      }
\label{fig:res_nus_spec} 
\end{figure}

\begin{figure*}
  \centering
    \includegraphics[scale=0.47]{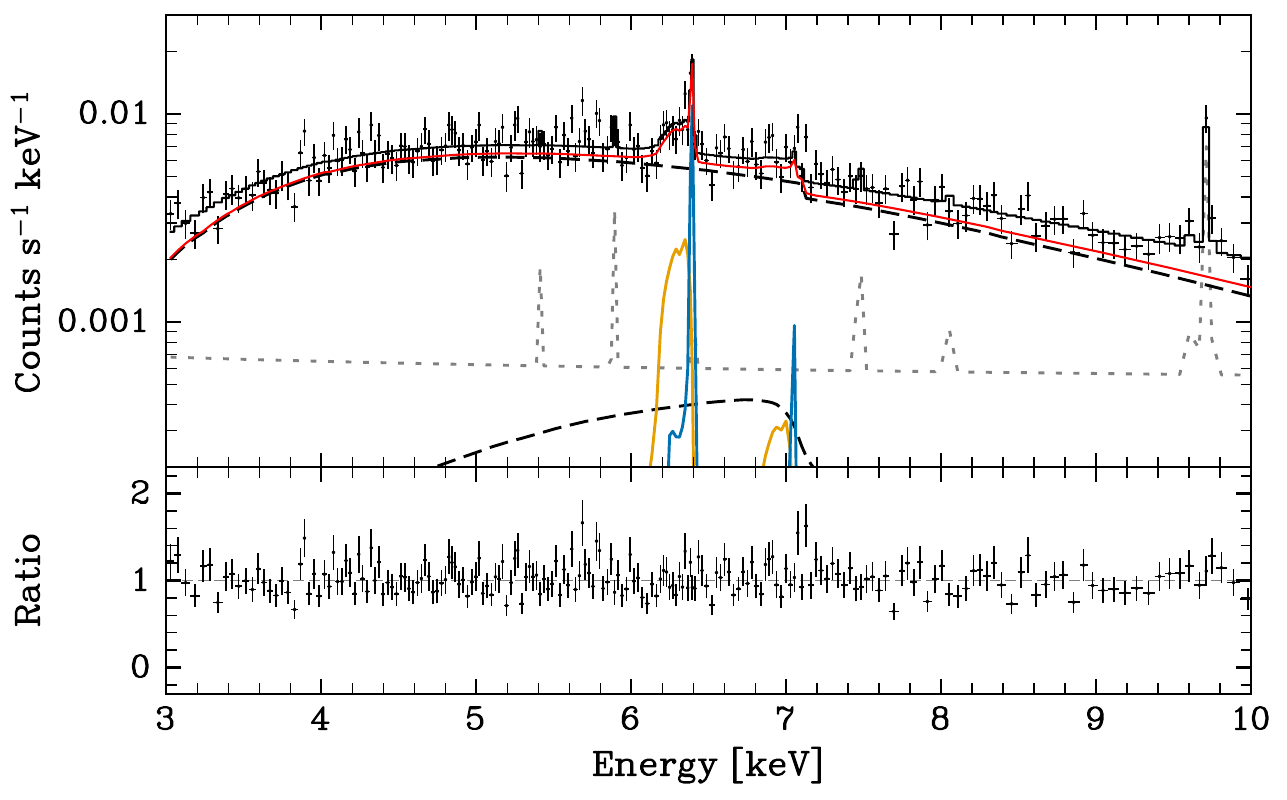}    
    \includegraphics[scale=0.47]{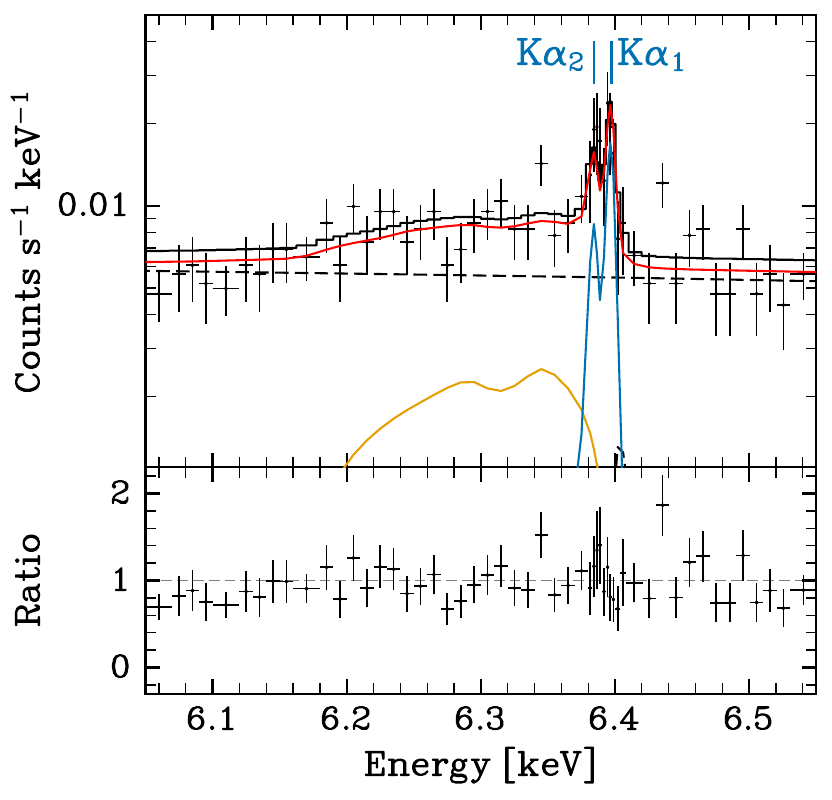}  
    \caption{
    \emph{Left:} Resolve spectrum (black crosses) and best-fit AGN model (red line).
    The broadened and narrow \texttt{MYTorusL} components are shown in orange and blue, respectively, and the powerlaw and reflection continuum components are 
    indicated by the black dashed lines. 
    The NXB model is shown by the dashed gray line. 
    \emph{Right:} Zoomed-in view of the Fe band. Finer binning is adopted to highlight the Fe~K$\alpha_1$ and Fe~K$\alpha_2$ components. In both figures, lower panels show data-to-model ratios.
      }\label{fig:first_res_spec} 
\end{figure*}

\subsubsection{Continuum Shape Determination Using NuSTAR}\label{sec:res_nus}

To establish a robust continuum for the Fe-band analysis, we jointly fitted the Resolve (2--12~keV) and NuSTAR (3--30~keV) spectra extracted during the $\sim$80~ks NuSTAR interval, excluding the 5.5--7.5~keV band. 
We first fitted an absorbed power-law model with cross-normalization, \texttt{constant*tbabs*ztbabs*zpowerlw} and an absorbed power-law component representing XRS1 contamination (Figure~\ref{fig:res_nus_spec}).  
The \texttt{tbabs} component accounts for Galactic absorption fixed at $N_{\rm H}^{\rm Gal}=4.34\times10^{20}$~cm$^{-2}$ \citep{HI4PI2016A&A...594A.116H}, while \texttt{ztbabs} represents additional absorption at the redshift of NGC~4395. 
The intrinsic absorbing column density ($N_{\rm H}$), photon index ($\Gamma$), and power-law normalization (Norm$_{\rm PL}$) were tied among the three spectra, whereas the \texttt{constant} factor was fixed to unity for Resolve and allowed to vary for FPMA and FPMB. 
The redshift was fixed at $z=0.001064$. 
This simple model yielded a relatively hard photon index, $\Gamma\simeq1.42$, compared with typical Seyfert continua \citep[e.g.,][]{Ric17bass}, plausibly because reflection and/or complex absorption were not included \citep[e.g.,][]{Moran2005AJ....129.2108M,Nardini2011MNRAS.417.2571N}. 
We therefore added a cold-reflection continuum using the \texttt{MYTorus} model \citep[hereafter \texttt{MYTorusC};][]{Mur09}, tying the photon index and normalization of the incident continuum to those of the primary power law. 
Because the reflector inclination had little effect on the fit, we fixed it at $37^\circ$, motivated by the inclination inferred for the molecular gas disk on a scale of $\sim$10~pc \citep{denBrok2015ApJ...809..101D}.  
When the reflector column density was allowed to vary, it was found to be close to $N_{\rm H}=10^{24}$~cm$^{-2}$, and we therefore fixed it at this value in the following fits. 
Although this addition lowered the AIC only marginally by $\Delta{\rm AIC}=-1.12$, it softened the photon index to $\Gamma\simeq1.55$. 
Furthermore, motivated by the complex absorption known in NGC~4395 \citep[e.g.,][]{Nardini2011MNRAS.417.2571N,Kammoun2019ApJ...886..145K}, we added a partial-covering absorber, leaving its column density ($N_{\rm H}^{\rm PC}$) and covering fraction (CF) free. 
The fitting result is shown in Figure~\ref{fig:res_nus_spec}. 
The partial-covering absorber did not significantly improve the AIC ($\Delta{\rm AIC}\simeq0$), but yielded $\Gamma=1.63^{+0.21}_{-0.14}$, consistent with previous broadband spectral modeling including complex absorption/reflection components \citep[e.g.,][]{Kammoun2019ApJ...886..145K}. 
Although this component is not statistically required by the AIC, we retained it
as a conservative phenomenological representation of the known complex
absorption in NGC~4395. 
This choice also allows absorption-related uncertainties to be propagated into the Fe~K analysis.
We therefore decided to  adopt the continuum model including \texttt{MYTorusC} and one partial-covering absorber as the baseline model for the subsequent Resolve Fe K spectral analysis.

\subsubsection{Time-averaged Resolve Spectrum} \label{sec:time_ave_res}

We investigated the Fe~K emission using the Resolve spectrum extracted from
the entire net exposure ($\approx$231\,ks). The continuum was described by the baseline model determined in Section~\ref{sec:res_nus}, and we here focus on the additional components
required in the Fe~K band. 
In the Resolve-only fits, the column density of the full-covering intrinsic
absorber, modeled with \texttt{ztbabs}, was fixed at
$N_{\rm H}=6.5\times10^{22}$~cm$^{-2}$, as determined from the joint
Resolve--NuSTAR continuum fit in Section~\ref{sec:res_nus}. 
This choice was made because this parameter is strongly degenerate with the
column density and covering fraction of the partial-covering absorber.

Figure~\ref{fig:first_res_spec} shows that the Fe~K$\alpha$ feature is composed of a narrow core and an excess extending to the red side. 
To quantify
the statistical requirement for each component, we constructed the model
step by step, as summarized in Table~\ref{tab:model_ladder}. In this table,
$\Delta C$ and $\Delta{\rm AIC}$ are evaluated relative to the corresponding
parent model to which the new component was added. 
For representative models, we list their best-fit parameters in
Table~\ref{tab:fitted_pars}.

We first fitted the spectrum with the baseline continuum alone (M0). The
addition of a narrow neutral Fe~K$\alpha$ component, modeled with
\texttt{zbfeklor} \citep{Hol_1997PhRvA..56.4554H}, significantly improved the fit
(M1; $\Delta{\rm AIC}=-45.80$). We then added the associated Fe~K$\beta$
line using \texttt{zbfekblor}, while leaving the 
normalization free and tying the
remaining parameters to those of K$\alpha$. 
This component also improved the fit (M2; $\Delta{\rm AIC}=-2.19$). 
These results confirm that the narrow neutral Fe~K$\alpha$ core and its associated K$\beta$
emission are required by the data.

Even after including these narrow lines, a redward excess remained around
the Fe~K$\alpha$ feature. As a phenomenological description, we added a broad
Gaussian component to M2, leaving its centroid energy, width, and normalization
free. This led to a clear improvement (M3; $\Delta{\rm AIC}=-20.15$), with a
redshifted energy of $6.324^{+0.036}_{-0.017}$\,keV. 
We then replaced this Gaussian with a
relativistic \texttt{diskline} component \citep{Fabian1989MNRAS.238..729F}. 
The rest energy
was fixed at 6.4~keV, the emissivity index was fixed at $\beta=-3$, and the
inclination angle, inner radius, and normalization were left free. 
Regarding $R_{\rm out}$, we
first examined a trial fit in which $R_{\rm out}$ was allowed to vary and obtained a broad 90\% confidence interval of
$R_{\rm out}\simeq80$--$2100\,R_{\rm g}$. 
As $R_{\rm out}$ was not robustly constrained,
we fixed it to two representative values,
$2\times10^2\,R_{\rm g}$ and $10^3\,R_{\rm g}$, denoted as M4a and M4b,
respectively. 
We adopted $2\times10^2\,R_{\rm g}$ rather than $10^2\,R_{\rm g}$ because, in some
time-resolved fits, $R_{\rm in}$ approaches $\sim100\,R_{\rm g}$. 
The latter value tests a more extended disk
scale motivated by previous UV/optical constraints 
\citep{McHardy2023MNRAS.519.3366M}.
Both diskline models
improved the fit relative to M2 by $\Delta{\rm AIC}\approx -23$ to $-22$ and
also gave slightly lower AIC values than the broad Gaussian model 
($\Delta$ AIC $= -3\sim-2$). Thus, the
redward excess can be naturally described as relativistically broadened Fe
K emission.

\begin{deluxetable*}{llrrrrrrr}
\tabletypesize{\scriptsize}
\tablecaption{Stepwise model comparison for the time-averaged Resolve  spectrum\label{tab:model_ladder}}
\tablewidth{0pt} 
\startdata \\
Model & Components & $C_{\rm obs}$ & $C_{\rm exp}$ & $\Delta C$ & d.o.f. &  $\Delta$d.o.f. & AIC & $\Delta$AIC \\ 
\hline
M0 & Continuum only 
   & 2079.55 & 2071.99$\pm$66.46 & ... & 1942 & ... & 2085.56 & ...  \\ 
       \multicolumn{9}{c}{\hdashrule[0.5ex]{0.95\textwidth}{0.4pt}{2pt 2pt}} \\
M1 & M0 + narrow Fe~K$\alpha$ (\texttt{zbfeklor}) 
   & 2029.73 & 2072.87$\pm$66.45 & $-$49.82 & 1940 & $-$2 & 2039.76 & $-$45.80 \\
M2 & M1 + narrow Fe~K$\beta$ (\texttt{zbfekblor}) 
   & 2025.53 & 2072.92$\pm$66.45 & $-$4.20 & 1939  & $-$1 & 2037.57 & $-$2.19 \\
M3 & M2 + broad Gaussian (\texttt{zgaussian}) 
   & 1999.33 & 2074.41$\pm$66.43 & $-$26.20 & 1936  & $-$3 & 2017.42 & $-$20.15 \\ 
M4a & M2 + diskline (\texttt{diskline}; $R_{\rm out} = 2\times10^2\,R_{\rm g}$) 
    & 1996.29 & 2074.33$\pm$66.43 & $-$29.24 & 1936  & $-$3 & 2014.38 & $-$23.19 \\ 
M4b & M2 + diskline (\texttt{diskline}; $R_{\rm out} = 10^3\,R_{\rm g}$) 
    & 1997.07 & 2074.19$\pm$66.43 & $-$28.46 & 1936  & $-$3 & 2015.16 & $-$22.41 \\ 
    \multicolumn{9}{c}{\hdashrule[0.5ex]{0.95\textwidth}{0.4pt}{2pt 2pt}} \\
M5 & M0 + blurred \texttt{MYTorusL} (\texttt{rdblur*MYTorusL}) 
   & 2014.14 & 2074.13$\pm$66.43 & $-$65.41 & 1939  & $-$3 & 2026.18 & $-$59.38 \\ 
M6 & M5 + BLR (\texttt{gsmooth*MYTorusL}) 
   & 2005.36 & 2074.36$\pm$66.42 & $-$8.78  & 1938  & $-$1 & 2019.42 & $-$6.76 \\ 
M7a & M5 + diskline (\texttt{rdblur*MYTorusL}; $R_{\rm out} = 2\times10^2\,R_{\rm g}$) 
   & 1998.57 & 2074.48$\pm$66.42 & $-$15.57 & 1937 & $-$2 & 2014.64 & $-$11.54 \\
M7b & M5 + diskline (\texttt{rdblur*MYTorusL}; $R_{\rm out} = 10^3\,R_{\rm g}$) 
   & 1998.38 & 2074.50$\pm$66.42 & $-$15.76 & 1937  & $-$2 & 2014.45 & $-$11.73 \\ 
    \multicolumn{9}{c}{\hdashrule[0.5ex]{0.95\textwidth}{0.4pt}{2pt 2pt}} \\ 
M5' & M0 + blurred \texttt{XClumpyL} (\texttt{rdblur*XClumpyL}) 
   & 2014.87 & 2074.34$\pm$66.42 & $-$64.68 & 1939  & $-$3 & 2026.91 & $-$58.65 \\ 
M6' & M5' + BLR (\texttt{gsmooth*XClumpyL}) 
   & 2006.47 & 2074.43$\pm$66.41 & $-$8.40  & 1938  & $-$1 & 2020.53 & $-$6.38 \\ 
M7a' & M5' + diskline (\texttt{rdblur*XClumpyL}; $R_{\rm out} = 2\times10^2\,R_{\rm g}$) 
   & 1997.71 & 2074.53$\pm$66.42 & $-$17.16 & 1937 & $-$2 & 2013.78 & $-$13.13 \\
M7b' & M5' + diskline (\texttt{rdblur*XClumpyL}; $R_{\rm out} = 10^3\,R_{\rm g}$) 
   & 1997.71 & 2074.53$\pm$66.42 & $-$17.16 & 1937 & $-$2 & 2013.78 & $-$13.13 \\    
\enddata
\tablecomments{
    $C_{\rm exp}$ and its variance were computed following
\citet{Kaa17}. $\Delta C$ and $\Delta{\rm AIC}$ are measured
relative to the parent model specified in the Components column. 
}
\end{deluxetable*}

\begin{deluxetable*}{cccccccccc}
\tabletypesize{\footnotesize}
\tablecaption{Best-fit parameters obtained for Resolve spectra averaged over the entire exposure\label{tab:fitted_pars}}
\tablewidth{0pt} 
\startdata \vspace{-0.1cm} \\ 
    &  Parameter & Units & M3 & M4a &  M4b & M6 & M7a & M7b \\ \hline 
(1) & $ N_{\rm H} $   & $10^{22}$ cm$^{-2}$         & 6.5$^{\rm f}$ & 6.5$^{\rm f}$ & 6.5$^{\rm f}$ & 6.5$^{\rm f}$ & 6.5$^{\rm f}$ & 6.5$^{\rm f}$ \\
(2) & $ N^{\rm PC}_{\rm H} $ & $10^{22}$ cm$^{-2}$ & $7.6^{+12.5}_{-4.9} $ & $ 7.3^{+12.1}_{-4.4} $ & $ 8.0^{+12.8}_{-4.9} $ & $7.0^{+15.6}_{-5.0}$ & $7.0^{+15.9}_{-5.0}$ & $ 6.7^{+16.0}_{-4.6} $  \\    
(3) & CF & ... & $0.48^{+0.42}_{-0.21}$ & $0.50^{+0.50}_{-0.22}$ & $0.48^{+0.51}_{-0.39}$ & $0.50^{+0.50}_{-0.24}$ & $0.50^{+0.50}_{-0.24}$ & $0.52^{+0.48}_{-0.25}$ \\
(4) & ${\rm Norm}_{\rm PL}$ & 10$^{-4}$ ph keV$^{-1}$ cm $^{-2}$ s$^{-1}$ &  $7.79^{+0.25}_{-0.11}$ & $7.81^{+0.44}_{-0.16}$ & $7.84\pm0.25$ & $7.76^{+0.50}_{-0.39} $ & $7.76^{+0.49}_{-0.37}$ & $7.75^{+0.49}_{-0.35}$  \\
(5) & $ \sigma_{\rm K\alpha} $ & km s$^{-1}$ & $ 0^{+107}  $ & $ 0^{+108} $ & $ 0^{+110}$ & ... & ... & ...      \\
(6) & $ {\rm Norm}_{\rm K\alpha} $  & 10$^{-6}$ ph cm $^{-2}$ s$^{-1}$ & $ 1.11^{+0.33}_{-0.42} $ & $ 1.14^{+0.44}_{-0.39} $ & $ 1.39 ^{+0.62}_{-0.52} $ & ... & ... & ...\\
(7) & $ {\rm Norm}_{\rm K\beta} $ & 10$^{-6}$ ph cm $^{-2}$ s$^{-1}$ & $ 0.37^{+0.34}_{-0.28} $ & $ 0.36^{+0.34}_{-0.28} $ &	$ 0.36^{+0.35}_{-0.27} $ & 	... & ... & ... \\
(8) & $ E_{\rm Gauss} $          &  keV & $ 6.324 ^{+0.036}_{-0.017}    $   & 	 ... & 	 ... &  	... & 	 ... &  	...    \\
(9) & $ \sigma_{\rm Gauss} $     &  keV & $ 0.072 ^{+0.028}_{-0.011} $   & 	 ... & 	 ... &  $0.067^{+0.079}_{-0.032} $ & 	 ...    \\
(10) & $ {\rm Norm}_{\rm Gauss} $ &  10$^{-6}$ ph cm $^{-2}$ s$^{-1}$ & $ 2.89^{+0.87}_{-1.06} $ & ... & 	 ... &  	... & 	 ... &  	...   \\
(11) & $ R^{\rm b}_{\rm in} $       &  $R_{\rm g}$ & ... &	$  69\pm11 $ & 	$ 60^{+15}_{-18} $ & 	... & $ 70^{+24}_{-15} $ & $ 68^{+31}_{-25} $    \\
(12) & $ R^{\rm b}_{\rm out} $      &  $R_{\rm g}$ & ... &	$  200^{\rm f}$ & 	$ 1000^{\rm f}$ &  	... & $200^{\rm f}$ & 	$ 1000^{\rm f}$ \\ 
(13) & $ \theta^{\rm b}_{\rm inc} $ &  deg & ... &	$  6.6^{+0.8}_{-1.0} $ & 	$4.7^{+3.2}_{-3.1}$ & 	... & $6.3^{+2.5}_{-1.9}$ & 	$ 6.5^{+1.9}_{-6.5} $    \\
(14) & $ {\rm Norm}^{\rm b}_{\rm DL} $ & 10$^{-6}$ ph cm $^{-2}$ s$^{-1}$ &  ... &	$ 2.66^{+0.93}_{-0.84} $ & $2.65^{+0.90}_{-0.86}$ & ... & ... & ...  \\
(15) & $ {\rm Norm}_{\rm MT}^{\rm b} $ & 10$^{-4}$ ph keV$^{-1}$ cm $^{-2}$ s$^{-1}$ & ... & ... & ... & $6.0^{+3.8}_{-2.9}$ & $5.5^{+2.4}_{-2.2}$ & $5.9^{+2.3}_{-2.5} $   \\
(16) & $ R^{\rm n}_{\rm in} $ & $10^6\,R_{\rm g}$ &  ... & ...  &  ...  & 	$9.9_{-9.3}$ & 	$9.9_{-6.3}$ & $9.9_{-9.3}$  \\
(17) & $ R^{\rm n}_{\rm out} $ & $10^6\,R_{\rm g}$ &  ...  & ...  &  ...  & 	$10^{\rm f}$ & $10^{\rm f}$ & $10^{\rm f}$     \\
(18) & $ \theta^{\rm n}_{\rm inc} $   & deg & ... & ... & ... & 	37$^{\rm f}$ & 37$^{\rm f}$ & 37$^{\rm f}$   \\
(19) & $ {\rm Norm}^{\rm n}_{\rm MT} $ & 10$^{-4}$ ph keV$^{-1}$ cm $^{-2}$ s$^{-1}$ & ... &	...  &  ...  & $2.8 ^{+1.2}_{-1.1}$	 & 	$3.1^{+1.1}_{-1.0}$ & 	$ 2.7 ^{+1.3}_{-1.0} $   \\
\enddata
\tablecomments{
The superscripts ``PC'', ``b'', and ``n'' denote the partial-covering absorber,
the broadened Fe~K component, and the narrow/distant \texttt{MYTorusL} component,
respectively. 
(1) Hydrogen column density of full-covering intrinsic absorber. 
(2)--(3) Hydrogen column density and covering fraction of the
partial-covering absorber. 
(4) Normalization of the primary power-law continuum at 1 keV.
(5)--(7) Velocity widths and normalizations of the
\texttt{zbfeklor} and \texttt{zbfekblor} components used in M3, M4a, and M4b
to represent the narrow neutral Fe~K$\alpha$ and K$\beta$ lines. 
(8)--(10) Phenomenological broad Gaussian component in M3; for M6, (9) represents the Gaussian smoothing width of the \texttt{gsmooth} component. 
(11)--(14) Inner radius, outer radius, inclination angle, and
normalization of the \texttt{diskline} component in M4a and M4b. 
For M7a and M7b, rows (11)--(13) give the corresponding parameters of the
broadened \texttt{rdblur*MYTorusL} component.
(15) Normalization of the broadened \texttt{MYTorusL} component.
(16)--(19) Inner radius, outer radius, inclination angle, and
normalization of the narrow/distant \texttt{MYTorusL} component used in M6, M7a, and M7b. 
The \texttt{MYTorusL} column density was fixed at $N_{\rm H}=10^{24}$ cm$^{-2}$, and
the incident photon index was tied to that of the primary power-law continuum.
Parameters marked with ``f'' were fixed during the fit. 
Uncertainties are quoted at the 90\% confidence level.
}
\end{deluxetable*}

\begin{deluxetable*}{cccccccccc}
\tabletypesize{\footnotesize}
\tablecaption{Best-fit parameters obtained with Model 7a for Resolve spectra produced in different time intervals\label{tab:time_res_fitted_pars}}
\tablewidth{0pt} 
\startdata \vspace{-0.1cm} \\ 
    &  Parameter & Units   & First half exp. & Second half exp. &  Low-flux time & High-flux time \\ \hline 
(1) & $ N^{\rm PC}_{\rm H} $                & $10^{22}$ cm$^{-2}$  & $ 1.8^{+46.1}_{-1.2} $ & $ 9.8^{+18.1}_{-7.2} $ & $ 3.1^{+29.6}_{-1.5}$ & $10.1^{+12.}_{-8.3}$  \\
(2) & CF                           &   ...         & 1.00$_{-0.92}$ & 0.49$^{+0.51}_{-0.19}$ & 1.00$_{-0.78}$ & 0.39$^{+0.61}_{-0.17}$ \\
(3) & $ {\rm Norm}_{\rm PL} $      & 10$^{-4}$ ph keV$^{-1}$ cm$^{-2}$ s$^{-1}$ & $ 5.92^{+0.57}_{-0.33}$ &	$ 9.50^{+0.83}_{-0.53} $ & $5.16^{+0.59}_{-0.32} $ & $ 10.26^{+0.84}_{-0.57} $ \\
(4) & $ R^{\rm b}_{\rm in} $             & $R_{\rm g}$ & $ 189^{+10}_{-70} $ & $ 145^{+54}_{-41} $ & $199_{-91} $ & 	$ 140^{+59}_{-60} $    \\
(5) & $ R^{\rm b}_{\rm out} $  & $R_{\rm g}$ & $ 200^{\rm f} $ & $ 200^{\rm f} $ & $ 200^{\rm f} $ & $ 200^{\rm f} $ \\
(6) & $ \theta^{\rm b}_{\rm inc} $       & deg         & $ 1.0^{+1.3} $ & 	$ 15.3^{+1.0}_{-1.3}$ & $ 1.0^{+2.0} $ & $ 14.4^{+1.9}_{-0.9} $   \\
(7) & $ {\rm Norm}^{\rm b}_{\rm MT} $       & 10$^{-4}$ ph keV$^{-1}$ cm$^{-2}$ s$^{-1}$ &	$ 2.7^{+1.6}_{-1.3} $ & $ 9.8^{+4.0}_{-3.8} $ & $ 2.7^{+1.6}_{-1.3} $ & $7.8^{+4.0}_{-3.6} $  \\
(8) & $ R^{\rm n}_{\rm in} $             & $10^6\,R_{\rm g}$ & 9.9$^{\rm f}$ & 9.9$^{\rm f}$ & 9.9$^{\rm f}$ & 9.9$^{\rm f}$ \\
(9) & $ R^{\rm n}_{\rm out} $            & $10^6\,R_{\rm g}$ & 10$^{\rm f}$ & 10$^{\rm f}$ & 10$^{\rm f}$ & 10$^{\rm f}$ 	\\
(10) & $ \theta^{\rm n}_{\rm inc} $      & deg         & 37$^{\rm f}$ & 37$^{\rm f}$ & 37$^{\rm f}$ & 37$^{\rm f}$ \\
(11) & ${\rm Norm}^{\rm n}_{\rm MT}$      & 10$^{-4}$ ph keV$^{-1}$ cm$^{-2}$ s$^{-1}$ & $3.2^{+1.4}_{-1.2}$ & $2.7^{+1.7}_{-1.4}$ & $3.8^{+1.5}_{-1.3}$ & $2.6^{+1.6}_{-1.4}$  \\ 
\enddata
\tablecomments{
The superscripts ``PC'', ``b'', and ``n'' denote the partial-covering
absorber, the broadened Fe~K component, and the narrow/distant MYTorus
component, respectively. 
The full-covering intrinsic absorber, modeled with \texttt{ztbabs}, was fixed
at $N_{\rm H}=6.5\times10^{22}$ cm$^{-2}$, as determined from the joint
Resolve--NuSTAR continuum fit. 
(1)--(2) Hydrogen column density and covering fraction of the
partial-covering absorber.
(3) Normalization of the primary
power-law continuum at 1 keV. 
(4)--(7) Inner radius, outer radius, inclination angle, and
normalization of the broadened \texttt{rdblur*MYTorusL} component. 
(8)--(11) Inner radius, outer radius, inclination angle, and normalization of the narrow/distant
\texttt{rdblur*MYTorusL} component. 
The \texttt{MYTorusL} column density was fixed at $N_{\rm H}=10^{24}$ cm$^{-2}$, and
the incident photon index was tied to that of the primary continuum. 
Parameters marked with ``f'' were fixed during the fit. 
}
\end{deluxetable*}

We also tested a more physically self-consistent description of the neutral
fluorescent emission using the \texttt{MYTorus} line component \citep[hereafter, \texttt{MYTorusL};][]{Mur09}. 
In this approach, the narrow Fe~K complex was modeled with \texttt{MYTorusL} convolved with \texttt{rdblur}. 
The column density of \texttt{MYTorusL} was fixed to $10^{24}$ cm$^{-2}$, while the inclination angle was left free. 
The emitting radius of \texttt{rdblur} was set to be large
enough for the component to remain unresolved ($R_{\rm out} = 10^7\,R_{\rm g}$), and we adopted $-3$ for the emissivity index. This model, M5, substantially improved the fit relative to the continuum-only model
($\Delta{\rm AIC}=-59.38$), demonstrating that the narrow Fe~K complex is also
well described by a physically motivated neutral-reflection model including
the Fe~K$\beta$ line and Compton-shoulder structure self-consistently.

\begin{figure*}
  \centering
\includegraphics[width=0.48\textwidth]{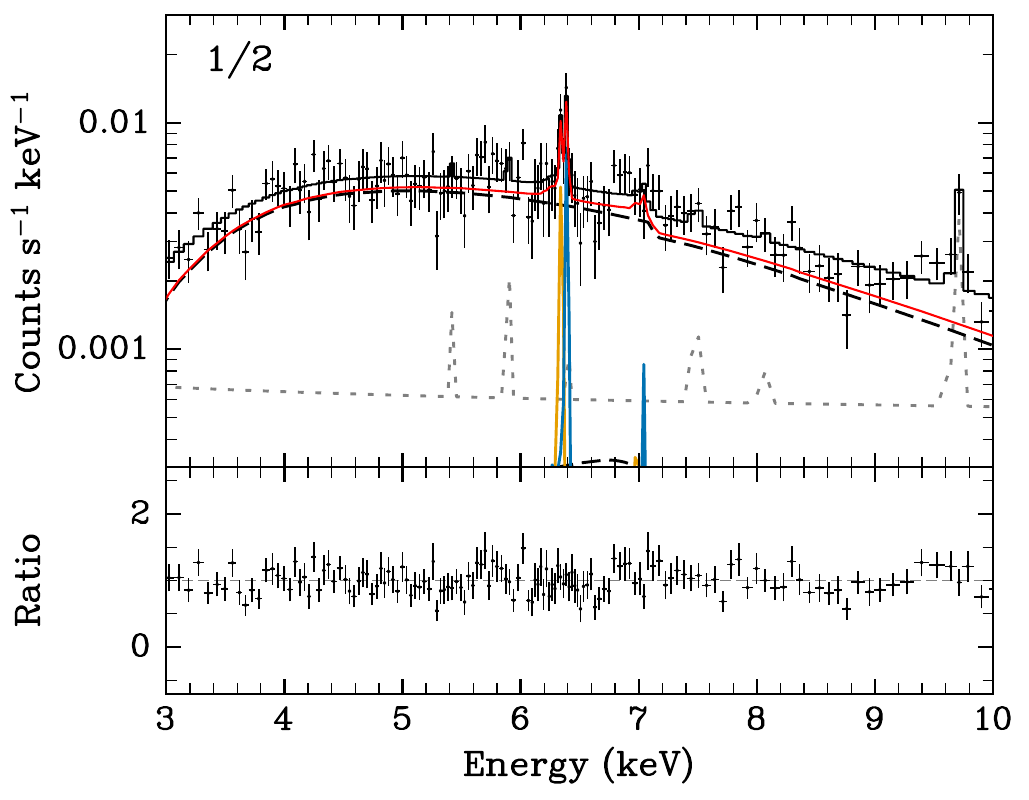}
\includegraphics[width=0.48\textwidth]{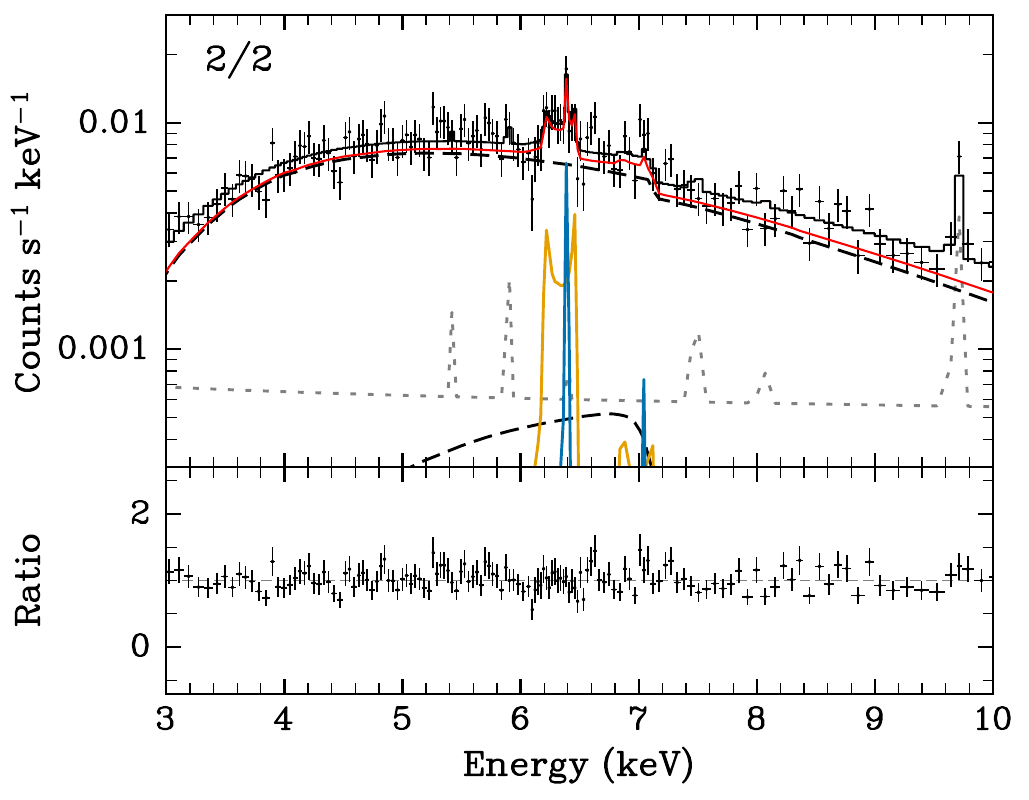}
    \caption{
    Resolve spectra from the first (left) and second (right) halves of the observation, 
    highlighting the time variation in the broadened Fe K profile.  
    The figures are shown in the same manner 
    as those in Figure~\ref{fig:first_res_spec}. 
  }\label{fig:spec_o2} 
\end{figure*}

We then examined whether the redward excess may instead be reproduced by
a modestly broadened, non-relativistic component, as might be expected from
the optical broad-line region. For this purpose, we added a
\texttt{gsmooth*MYTorusL} component to M5. This model improved the AIC
(M6; $\Delta{\rm AIC}=-6.76$), but the required broadening corresponds to a
velocity width ($\sim 3100$ km s$^{-1}$) 
much larger than that of the optical H$\alpha$ broad component
reported for NGC~4395 \citep[e.g., 426 km s$^{-1}$;][]{Woo2019NatAs...3..755W}. We therefore do not regard the BLR interpretation as
the most natural explanation of the redward Fe~K excess.

Finally, we modeled the redward excess with a relativistically blurred
\texttt{MYTorus} fluorescent-line component, \texttt{rdblur*MYTorusL}, added to the
narrow \texttt{MYTorusL} component. 
Following the same strategy as for the \texttt{diskline} fits, we fixed the outer
radius of the broadened \texttt{rdblur*MYTorusL} component to
$R_{\rm out}=2\times10^2\,R_{\rm g}$ and $10^3\,R_{\rm g}$, corresponding to
M7a and M7b, respectively.
The inclination angle of the narrow \texttt{MYTorusL} component was fixed at $37^\circ$, motivated by the
parsec-scale molecular gas disk \citep{denBrok2015ApJ...809..101D}. 
Both models improved the fit relative to M5
($\Delta{\rm AIC}=-11.54$ and $-11.73$), indicating that an additional
relativistically broadened Fe~K component is statistically favored.
The
two choices of $R_{\rm out}$ gave nearly identical AIC values and consistent
best-fit parameters (Table~\ref{tab:fitted_pars}). We hereafter adopt M7a as the baseline model for the
time-averaged Fe~K spectrum, while confirming that the choice between
$R_{\rm out}=2\times10^2\,R_{\rm g}$ and $10^3\,R_{\rm g}$ does not affect the
main conclusions.

For the adopted model M7a, the measured $C$ statistic is
$C_{\rm obs}=1998.57$, while the expected value is
$C_{\rm exp}=2074.48\pm66.42$, corresponding to
$C_{\rm obs}/C_{\rm exp} \approx 0.96$ and 
$(C_{\rm obs}-C_{\rm exp})/\sqrt{C_{\rm v}}\approx -1.1$.
The final time-averaged fit is consistent with the expected $C$-statistic distribution
for the adopted binning (see the results for \texttt{opt1} in Figure~\ref{fig:c_vs_binning} in Appendix A).

As an important check on the modeling of the Fe K$\alpha$ red wing,
we examined whether our results depend on the assumed treatment of the
scattering electrons in the fluorescent-line profile. 
In particular, we tested whether replacing \texttt{MYTorusL} with 
the \texttt{XClumpy} torus model, which treats the Compton shoulder 
from bound electrons \citep{Tan19,XRISM2026NatAs.tmp...63X}, 
affects the inferred origin of the redward excess. 
For this purpose, we repeated the fits corresponding to Models M5, M6,
M7a, and M7b by replacing \texttt{MYTorusL} with
the \texttt{XClumpy} fluorescent-line component (\texttt{XClumpyL}). 
The parameters common to
\texttt{MYTorusL} and \texttt{XClumpyL}
were set in the same way as in the MYTorus-based
fits, and the additional \texttt{XClumpy} angular-width parameter was initially
fixed at $30^\circ$.
The resulting fit statistics were comparable to those obtained with
\texttt{MYTorusL} (Table~\ref{tab:model_ladder}). 
We also confirmed that varying the angular width over \(10^\circ\)--\(60^\circ\) does not affect the results.
We therefore find that the use of a bound-electron Compton-shoulder profile does not significantly alter our interpretation of the time-averaged Fe K$\alpha$ red wing.

\subsubsection{Time- and Flux-resolved Resolve Spectra: Two-bin Case}\label{sec:twobin}

Motivated by the significant variability in the Resolve light curve
(Figure~\ref{fig:xlc}), we examined whether the Fe-band components varied.
We first divided the Resolve exposure into two equal
time segments and fitted the two spectra independently with the final 
model (M7a) determined in Section~\ref{sec:time_ave_res}. 
In these time-resolved fits, the inner radius of the narrowest broadened
component was fixed to its time-averaged best-fit value because it was not
well constrained in the individual time segments.
The fitted spectra are shown in Figure~\ref{fig:spec_o2}, and 
the resulting parameters are listed in Table~\ref{tab:time_res_fitted_pars}.
The inclination angle significantly changes from
$\theta_{\rm inc} < 2.3$~deg in the first half to
$\theta_{\rm inc}=15.3^{+1.0}_{-1.3}$~deg in the second half, with the
90\% confidence intervals not overlapping. 

We also performed a flux-resolved analysis by extracting two Resolve spectra from low- and high-flux intervals separated by the median count rate in the 2--10\,keV light curve (Figure~\ref{fig:xlc}b).
Fitting them with the same \texttt{rdblur*MYTorusL}-based model suggests
changes in the power-law normalization and in the broadened Fe-K component
(Table~\ref{tab:time_res_fitted_pars}).
In particular, the inclination angle of the broadened component changes
between the low- and high-flux intervals.
In principle, a more detailed flux-resolved analysis could be performed by
defining multiple count-rate thresholds. However, given the limited contrast between finer flux bins, we do not pursue a more subdivided flux-resolved analysis in this paper and instead focus on the time-resolved results.

\subsubsection{Time-resolved Resolve Spectra with Finer  Binnings}\label{sec:finebin}

\begin{figure}
  \centering
    \includegraphics[scale=0.353]{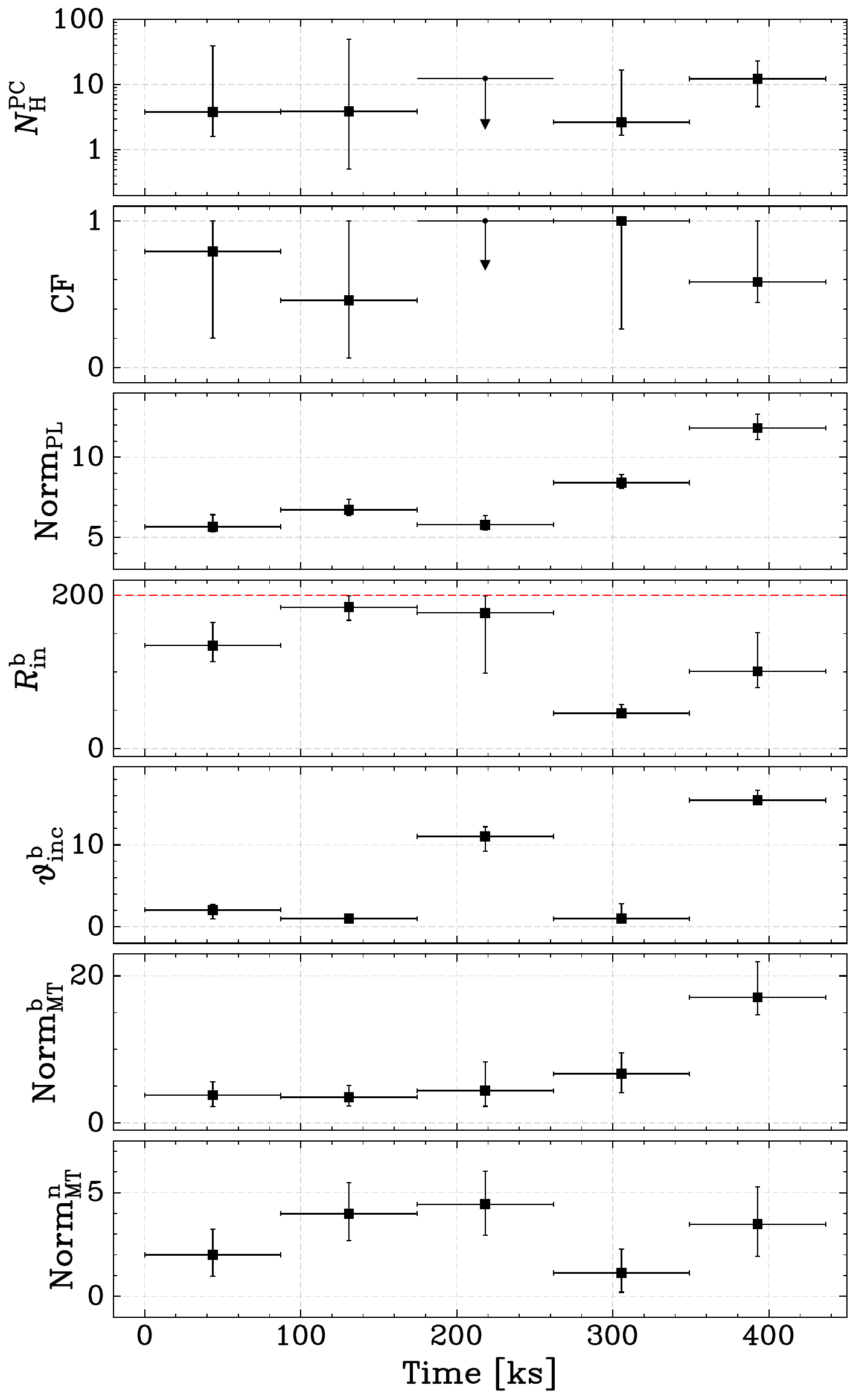}
  \caption{
    Time evolution of best-fit parameters obtained from five time-resolved spectra. 
    From top to bottom, the parameters are 
    the hydrogen column density ($10^{22}$ cm$^{-2}$) and covering fraction of the partial absorber, the power-law normalization ($10^{-4}$ photons keV$^{-1}$ cm$^{-2}$ s$^{-1}$ at 1\,keV), the inner radius ($R_{\rm g}$), 
    inclination angle (deg), and normalization ($10^{-4}$ photons keV$^{-1}$ cm$^{-2}$ s$^{-1}$) of the broad-line component, and the normalization of the narrow-line component
    ($10^{-4}$ photons keV$^{-1}$ cm$^{-2}$ s$^{-1}$). 
    The red dashed line in the $R_{\rm in}$ panel marks the fixed outer radius ($R_{\rm out}=200\,R_{\rm g}$).
  }\label{fig:five_bin_par_changes} 
\end{figure}

As the diskline emission modeled with \texttt{rdblur*MYTorusL} was found to have varied significantly during the XRISM observation (Section~\ref{sec:twobin}), we then examined the variability on a shorter timescale by dividing the Resolve exposure into five time segments.
The fixed and free parameters were the same as in the two-bin analysis. 
The resultant time evolutions of fitted parameters are shown in Figure~\ref{fig:five_bin_par_changes}.

To quantify the variability, we fitted the five measurements of
$R_{\rm in}$ and $\theta_{\rm inc}$ in Figure~\ref{fig:five_bin_par_changes} with constant models, accounting for the asymmetric statistical uncertainties. The constant models are rejected with null-hypothesis probabilities of
$p \ll 0.01$ for both $R_{\rm in}$ and $\theta_{\rm inc}$. In contrast, the
normalization of the narrow Fe~K$\alpha$ component is consistent with a
constant value ($p \sim 0.3$). These results indicate that the observed time variability is mainly associated with the broadened Fe~K component, while the distant narrow fluorescent emission remains stable within the statistical uncertainties.

As an alternative interpretation, we also tested whether the observed
time-dependent profile changes could be described by changes in the column
density of the line-emitting material rather than by changes in the geometric
parameters. For this test, we fixed the $R_{\rm in}$ and $\theta_{\rm inc}$ of the
broadened \texttt{rdblur*MYTorusL} component to their 
time-averaged values and allowed the \texttt{MYTorusL} column density of this component
to vary among the five time segments. This model did not improve the fit
relative to the corresponding fixed-column-density model
($\Delta{\rm AIC}\simeq 0$ or positive). The column densities were also poorly
constrained.
As a complementary check, we repeated this test using \texttt{XClumpyL},
which treats Compton scattering from bound electrons. We allowed
$N_{\rm H}$ and the torus angular width, both of which affect the Compton-shoulder profile \citep{Tan19}, to vary among the time segments, 
and confirmed that this modification did not significantly improve the fit. 
We therefore interpret the profile evolution mainly in terms of changes in 
the broadened-component geometry, rather than variations in the 
line-emitting/scattering material or in the bound-electron Compton-shoulder profile.

\begin{figure*}
  \centering 
    \includegraphics[width=\textwidth]{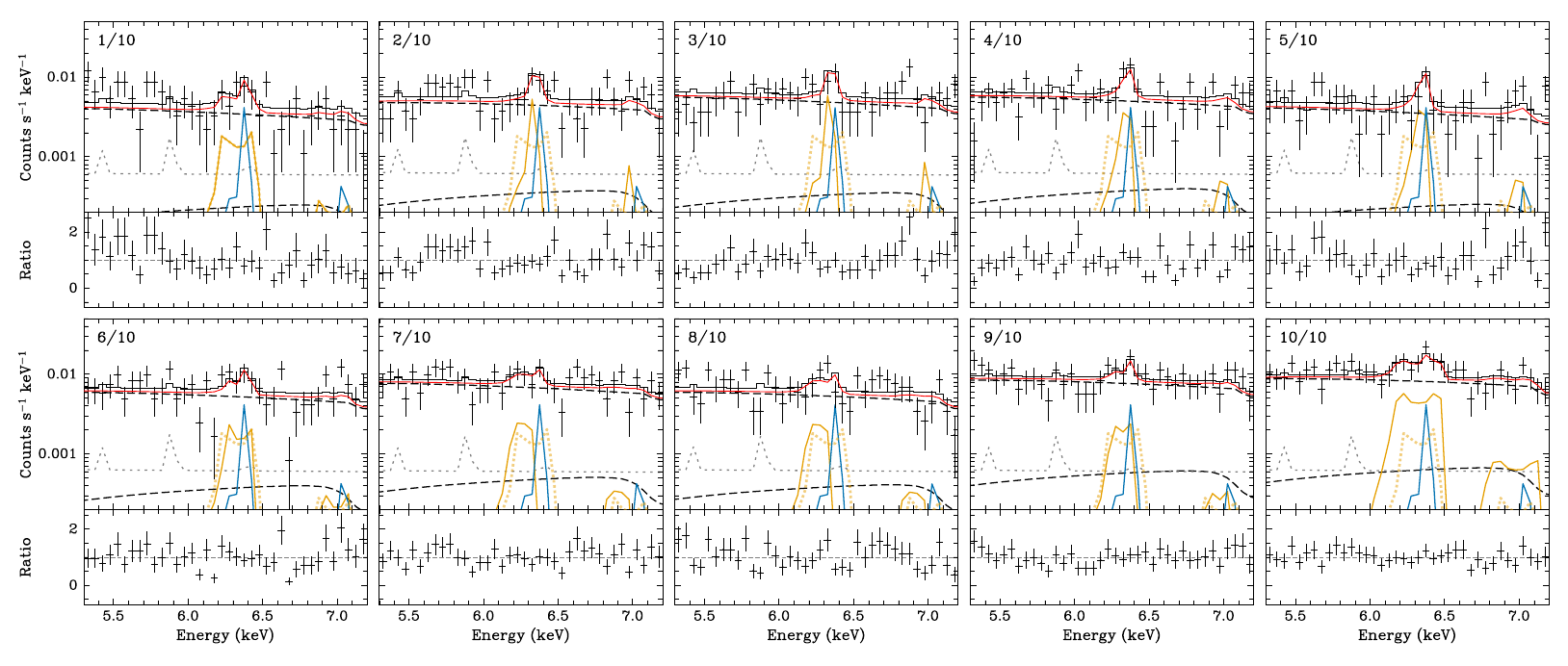}
      \vspace{-0.8cm}  
  \caption{
    Ten time-resolved Resolve spectra obtained by dividing the exposure into 10 segments. 
    The figures are shown in the same manner
    as Figure~\ref{fig:first_res_spec}, except for the additional dotted orange lines, indicating the broad \texttt{rdblur*MYTorusL} profile constrained in the first bin for comparison.  
    Residual line-like features in individual panels were examined with the
    look-elsewhere (Section~\ref{sec:finebin}); none required an additional Gaussian component
    at the 90\% confidence level.
  }\label{fig:10spec}  
\end{figure*}

Finally, motivated by the significant variation found in the five-bin
analysis, we extracted 10 time-resolved spectra. Because the photon statistics
in each ten-bin spectrum are limited, we fixed $R_{\rm in}$ in each ten-bin
fit to the best-fit value obtained from the corresponding five-bin spectrum
that covers the same time interval. We then allowed only $\theta_{\rm inc}$, the two partial absorber parameters, and the continuum normalization to vary. 
Exceptionally in the last bin, because allowing Norm$_{\rm disk}$ to vary significantly improved the fit, we kept it free. 
The ten-bin analysis is therefore not used to prove the existence of
variability by itself; rather, it is used to examine the possible timescale of
the inclination modulation suggested by the higher-statistics 
five-bin analyses. The fit results are shown in Figure~\ref{fig:10spec}.

\begin{figure}
  \centering 
    \includegraphics[width=0.45\textwidth]{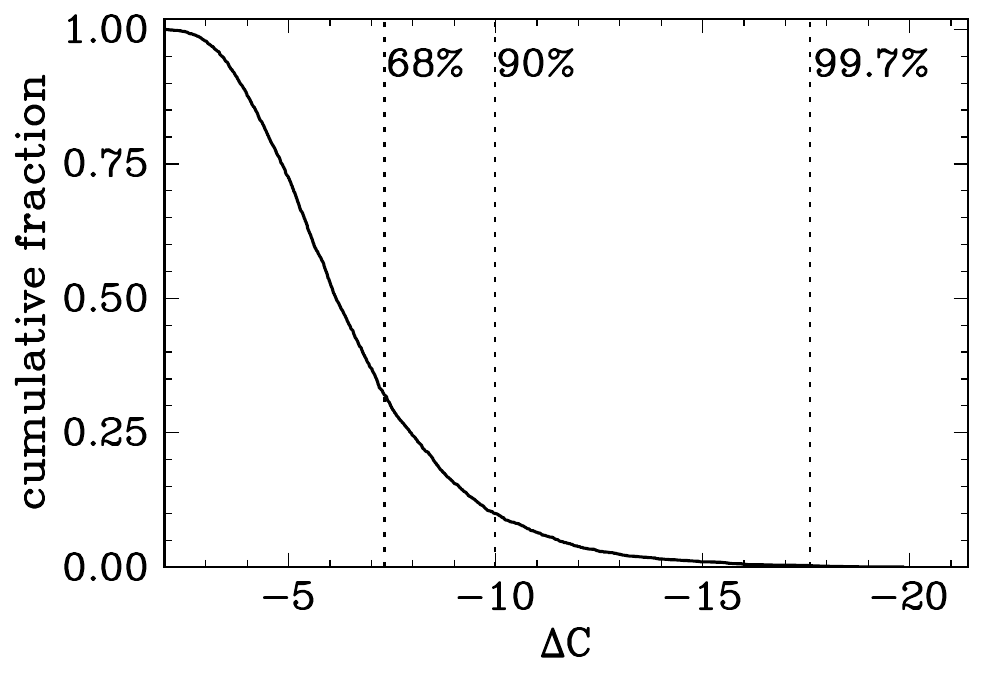}
      \vspace{-0.3cm}  
  \caption{
    Null distribution of the maximum line-search improvement obtained from the
Monte Carlo simulations. The distribution accounts for the look-elsewhere effect over both
trial line energy and 10 epochs. The vertical dashed lines indicate the 68\%,
90\%, and 99.7\% thresholds, corresponding to
$\Delta C=-7.3$, $-10.0$, and $-17.6$, respectively.
  }\label{fig:null_dist}  
\end{figure}

\begin{figure}
  \centering
  \includegraphics[scale=0.55]{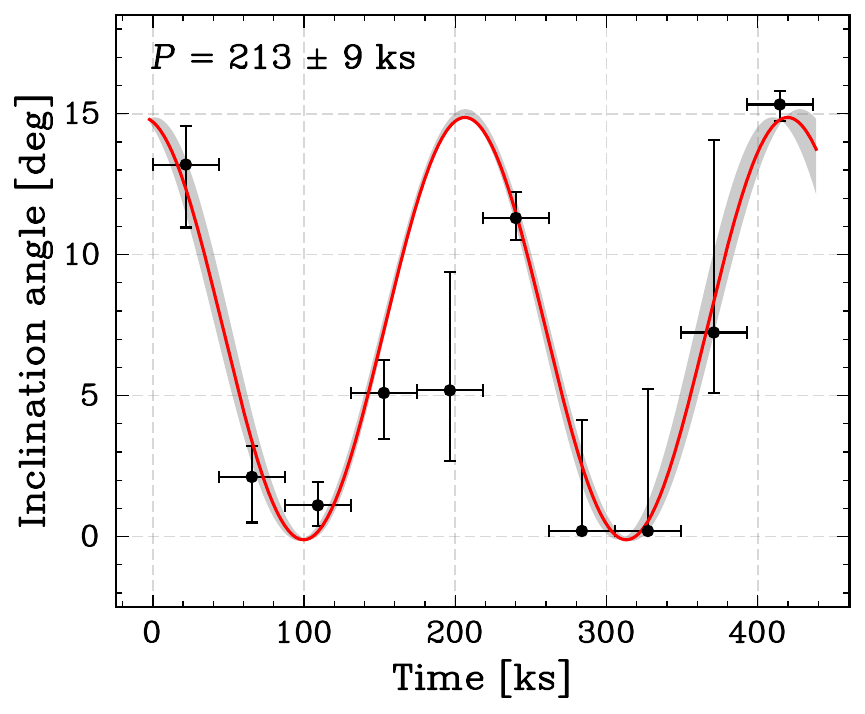}
    \caption{
    Inclination-angle evolution inferred from the 10 time-resolved Resolve spectra. 
    Error bars show 1$\sigma$ statistical uncertainties. 
    The red curve is the best-fit sinusoidal curve  
    with a period of 213\,ks, and 
    the gray band reflects the error in the period. 
      }\label{fig:ch_in_inc} 
\end{figure}

Before interpreting the time evolution of the broadened component, we tested
whether the residuals seen in the individual ten-bin spectra require additional
line-like components. For this purpose, we defined the line-search procedure as follows.
For each time-resolved spectrum, the corresponding best-fit model was adopted
as the null model, and an additional Gaussian line was scanned over
5.2--7.2~keV.
The Gaussian width was fixed at $\sigma=0.07$~keV, motivated by the broad
Gaussian width obtained from the time-averaged Fe-band fit (M3 in Table~\ref{tab:fitted_pars}).
The line centroid was scanned over 57 grid points separated by 0.035~keV; at
each grid point, only the line normalization was allowed to vary.
We defined the line-search improvement as
$\Delta C=C_{\rm line}-C_{\rm null}$ and used the most negative
$\Delta C$ value in each spectrum as the maximum improvement.

We first calibrated the significance of this search, including the
look-elsewhere effect, by using Monte Carlo simulations.
For each of the 10 epochs, we generated 3000 fake spectra from the
corresponding null model, using the same responses, background setup,
exposure, and spectral grouping as the real data, and applied the defined 
line-search procedure.
The maximum improvements from all 10 epochs were then combined to construct
the null distribution (Figure~\ref{fig:null_dist}). 
This distribution gives thresholds of
$\Delta C=-7.3$, $-10.0$, and $-17.6$ at the 68\%, 90\%, and 99.7\%
levels, respectively.
We then applied the same line-search procedure to the real ten-bin spectra.
No residual feature showed an improvement larger than the 90\% threshold
($\Delta C<-10.0$).
We therefore did not include extra Gaussian components in the time-resolved
fits.

To quantify the significance of the broadened \texttt{rdblur*MYTorusL} emission, we simply compared the fits with and without the component using AIC. Although the component is not individually required in every time bin, it is favored with
$\Delta{\rm AIC} \sim -3$ in epochs 2/10, 4/10, 6/10, and 8/10, and is strongly
favored in epoch 10/10 with $\Delta{\rm AIC}\simeq -16$. 
We further confirmed that the inferred time variation is not driven only by
the least significant epochs. Even when the constant-model test is restricted
to the epochs in which the broadened component is favored by AIC, the
inclination-angle measurements are inconsistent with a constant value
($p \ll 10^{-3}$).

Since Figure~\ref{fig:ch_in_inc} may suggest that the inclination angle shows a possible periodic trend, 
we fitted the measurements with a sinusoidal function of $A \sin (2\pi t / P + B) + C$ and determined best-fit values of $A$, $P$, $B$, and $C$ by minimizing $\chi^{2}$ 
while accounting for asymmetric uncertainties. 
We obtained a period of $213\pm 9$~ks, where the uncertainty was estimated from 
$\Delta\chi^{2}=2.706$ by stepping $P$ away from its best-fit value.\\

\section{Discussion: Time-variable Diskline} \label{sec:dis}\label{sec:time_var_broad}

\begin{figure}
  \centering
    \includegraphics[scale=0.42]{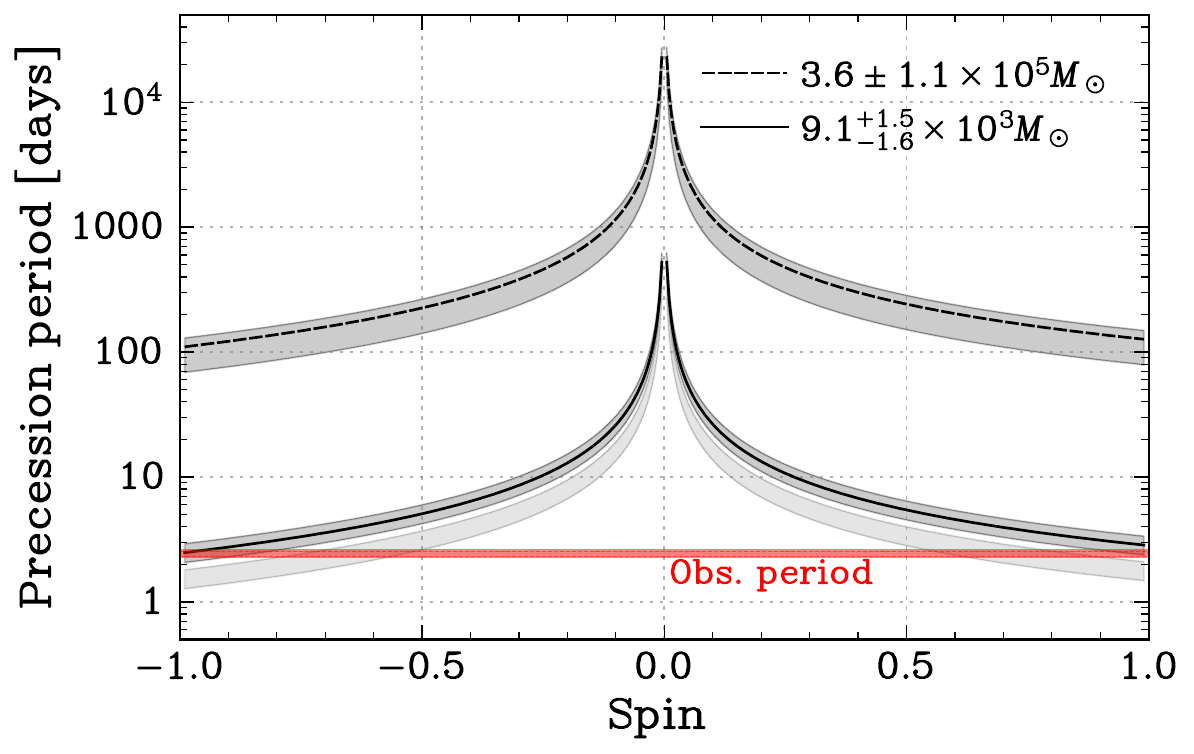}
    \caption{
    Observed modulation period of $213\pm9$ ks (red horizontal line) compared
    with LT-precession periods predicted by the rigid-disk model of
    \citet{Franchini2016MNRAS.455.1946F} as a function of BH spin. Curves with shaded dark bands show
    the predictions for two BH-mass assumptions,
    $M_{\rm BH}=9.1^{+1.5}_{-1.6}\times10^3M_\odot$ and
    $3.6\pm1.1\times10^5M_\odot$, using
    $R_{\rm in}=100\,R_{\rm g}$ and $R_{\rm out}=1.3\times10^2\,R_{\rm g}$.
    Here, $R_{\rm in}$ is the representative value obtained from a constant fit to
    the five-bin $R_{\rm in}$ measurements, while $R_{\rm out}$ is adopted as the
    largest outer radius that remains consistent with the observed period for the
    low-mass case. 
    The shaded lighter band indicate the extreme compact configulation with $R_{\rm out}$ only slightly larger than $R_{\rm in}$ for the lower mass case.
    }
\label{fig:lt}  
\end{figure}

The XRISM/Resolve spectrum of NGC~4395 in Figure~\ref{fig:first_res_spec}
reveals a narrow neutral Fe~K$\alpha$ core accompanied by a redward wing that can be well described by a relativistically broadened line
(Section~\ref{sec:time_ave_res}). 
Time-sliced spectroscopy further indicates that the broadened component varied
significantly within the $\sim400$~ks observation
(Sections~\ref{sec:twobin}--\ref{sec:finebin}).
Although $R_{\rm in}$ and $\theta_{\rm inc}$ were allowed to vary
simultaneously in the time-resolved fits, their physical implications are
discussed separately below for clarity.

\subsection{Inward motion of the inner disk region}

The five-bin analysis suggests a decrease in the inner radius of the
line-emitting region with time (Figure~\ref{fig:five_bin_par_changes}),
possibly indicating inward motion of the inner disk. 
At a representative radius of $R\sim\,100\,R_{\rm g}$, obtained from a constant fit to the five-bin $R_{\rm in}$ measurements, the viscous timescale
$t_{\rm vis}\sim\alpha^{-1}(H/R)^{-2}\Omega^{-1}$ is
$\sim500$--5000~ks for $M_{\rm BH}\sim10^{4-5}\,M_\odot$,
$\alpha\sim0.01$, and $H/R\sim0.1$. 
This range is comparable to or longer than the observation duration, but still would 
allow appreciable radial evolution. 
The contemporaneous X-ray brightening may also be consistent with this
scenario, since an inward-moving disk could increase the seed-photon supply to
the corona and enhance the Comptonized power-law emission.

\subsection{Periodic inclination changes and possible LT precession}\label{sec:lt_inter}

The ten-bin analysis suggests that the inclination varied with time and can be described by a sinusoid with a period of $P= 213\pm9$~ks (Figure~\ref{fig:ch_in_inc}).
This behavior is suggestive of geometric modulation. 
A natural possibility is LT precession of a tilted inner disk about the BH spin axis, which would modulate the apparent inclination inferred from the line profile.

Assuming that the modulation in $\theta_{\rm inc}$ is driven by LT
precession, we compare the observed period with theoretical expectations
following the rigid-disk precession model of \citet{Franchini2016MNRAS.455.1946F}. The
global precession frequency is computed by weighting the local LT torque by
the disk angular momentum,
\[
\Omega_{\rm p} =
\frac{
\int_{R_{\rm in}}^{R_{\rm out}}
\Omega_{\rm LT}(R)L(R)\,2\pi R\,dR
}{
\int_{R_{\rm in}}^{R_{\rm out}}
L(R)\,2\pi R\,dR
},
\]
where $\Omega_{\rm LT}(R)$ is the local LT precession frequency and $L(R)$ is
the angular-momentum surface density. At $r\equiv R/R_{\rm g}\gg1$,
$\Omega_{\rm LT}(R)\simeq 2a(c^3/GM_{\rm BH})r^{-3}$, where $a$, $c$, and
$G$ are the BH spin parameter, the speed of light, and the gravitational
constant, respectively.

For this comparison, we need to specify the characteristic radial range of the precessing line-emitting region. We adopted $R_{\rm in}=100\,R_{\rm g}$, obtained by fitting the five-bin measurements of $R_{\rm in}$ with a constant model. This value is used only as a representative inner radius for the LT-precession calculation.
For the outer radius, we adopted $R_{\rm out}=1.3\times10^2\,R_{\rm g}$. This value is not a directly measured spectral parameter. Instead, it is chosen as
the largest outer radius for which the predicted LT-precession period can
remain consistent with the observed period for the low-BH-mass case within the
allowed spin range. Since increasing $R_{\rm out}$ adds more slowly precessing
outer material and lengthens the global precession period, larger values of
$R_{\rm out}$  make it difficult to reproduce the observed modulation
period. 
Thus, the LT comparison below should be regarded as a fiducial interpretation for a compact precessing region with $R\simeq100$--$130\,R_{\rm g}$.

With this fiducial radial range, we compare the observed period with the two
plausible extremes of the BH mass estimates:
$M_{\rm BH}=9.1^{+1.5}_{-1.6}\times10^3\,M_\odot$ and
$3.6\pm1.1\times10^5\,M_\odot$
\citep[][]{Peterson2005ApJ...632..799P,Woo2019NatAs...3..755W}.
The lower and higher mass estimates are based on H$\alpha$ and C\,{\sc iv}
reverberation mapping, respectively.
As shown in Figure~\ref{fig:lt}, the observed period is most readily
reproduced for the low-mass case, because the LT-precession period scales
approximately with $M_{\rm BH}$ and becomes too long for the high-mass case.
For the low-mass case, the observed period requires a high spin of 
$a\sim0.9$. 
If we adopt an extreme compact
configuration with $R_{\rm out}$ only slightly larger than
$R_{\rm in}\simeq100\,R_{\rm g}$, a prograde spin of $a\gtrsim0.6$ is 
required.
Thus, if the inclination modulation is interpreted as LT
precession, the data favor the lower BH-mass estimate, a moderate spin ($a \gtrsim 0.6$), and a compact precessing region at $R\simeq100$--$130\,R_{\rm g}$.

The alternative spectral model M7b, in which the outer radius of the
\texttt{rdblur*MYTorusL} component is fixed at $10^3\,R_{\rm g}$, gives the same
qualitative time-resolved spectral results. However, the outer radius should be
distinguished from the LT-precession calculation: if the entire region out to
$10^3\,R_{\rm g}$ were assumed to precess rigidly, the predicted period would become substantially longer. The LT interpretation therefore requires that the
effective precessing region be compact, even if the spectral model allows a larger formal outer radius.

Although our baseline model uses \texttt{rdblur}, which assumes a
Schwarzschild metric, the LT interpretation implies a spinning BH.
As a consistency check, we replaced \texttt{rdblur} in the model M7a with the Kerr convolution kernel \texttt{kerrconv} \citep{Brenneman2006ApJ...652.1028B} and refitted
the time-averaged Resolve spectrum.
Because a full exploration of the \texttt{kerrconv}-based model is
computationally expensive, we
fixed the spin at 0.90 and the inclination angle at the best-fit value
obtained with the baseline \texttt{rdblur}-based model.
Regarding the broadened component, 
only the inner radius and normalization were left free.
The Kerr-convolved model reproduced the observed redward Fe-K profile,
yielding $R_{\rm in}=77^{+29}_{-24}\,R_{\rm g}$, consistent with the
characteristic radius inferred from the baseline fit.
Thus, our conclusions would not be strongly sensitive to the choice of a
Schwarzschild or Kerr broadening kernel.

A key requirement for the LT interpretation is that the tilted region precesses approximately coherently rather than being rapidly twisted by differential precession \citep[e.g.,][]{Fragile2024arXiv240410052F}. 
For the verification, there are two commonly used criteria. 
The first criterion concerns whether tilts, or warps, can propagate fast enough to maintain coherence across the precessing region.
Given that a warp is carried by pressure forces at roughly the sound speed,  the relevant sound-crossing time should be shorter than the LT precession time, so that near-rigid-body precession becomes plausible \citep[e.g.,][]{Papaloizou1995MNRAS.274..987P}. 
The second criterion concerns viscous alignment. 
When the disk is geometrically thick enough compared to the viscosity ($H/R \gtrsim \alpha$), viscous alignment is less efficient, thereby mitigating the Bardeen--Petterson effect \citep{Bardeen1975ApJ...195L..65B}. As a result, disk warps can propagate over a wide radial range.

    We quantify the first condition by comparing the azimuthal sound-crossing
time, $\tau_{\rm cs}\equiv 2\pi R/c_s$, with the local LT precession period,
$\tau_{\rm LT}\equiv 2\pi/\Omega_{\rm LT}$, following
\citet{Fragile2005ApJ...623..347F}.
For a gas-pressure-dominated fluid, the adiabatic sound speed satisfies
$c_s^2=\Gamma P/\rho$, where $\Gamma$, $P$, and $\rho$ are the adiabatic
index, pressure, and density, respectively.
We approximate
$c_s\simeq \sqrt{\Gamma}\,(H/R)\,v_{\rm K}$, where
$v_{\rm K}=(GM/R)^{1/2}=c\,r^{-1/2}$ and $r\equiv R/R_{\rm g}$.
This gives
\[
\tau_{\rm cs}=
\frac{2\pi GM_{\rm BH}}{c^3}
\frac{r^{3/2}}{\sqrt{\Gamma}\,(H/R)} .
\]
Using
$\Omega_{\rm LT}\simeq 2a(c^3/GM_{\rm BH})r^{-3}$ for $r\gg1$, the LT
precession period is
\[
\tau_{\rm LT}=
\frac{\pi GM_{\rm BH}}{a c^3}r^3 .
\]
Equating $\tau_{\rm cs}$ and $\tau_{\rm LT}$ yields the characteristic global precession radius
\begin{eqnarray}
r_{\rm gp}&\equiv& \frac{R_{\rm gp}}{R_{\rm g}}
=\left[\frac{2a}{(H/R)\sqrt{\Gamma}}\right]^{2/3} \nonumber \\ 
&\simeq& 26.9
\left(\frac{a}{0.9}\right)^{2/3}
\left(\frac{H/R}{0.01}\right)^{-2/3}
\left(\frac{\Gamma}{5/3}\right)^{-1/3}.
\end{eqnarray}
Adopting $H/R\gtrsim0.01$, consistent with the second criterion
$H/R\gtrsim\alpha$ for $\alpha\sim0.01$, and $\Gamma=5/3$, we obtain
$R_{\rm gp}\lesssim27\,R_{\rm g}$.
Thus, the fiducial precessing region adopted in Figure~\ref{fig:lt},
$R\simeq100$--$130\,R_{\rm g}$, lies outside $R_{\rm gp}$ and satisfies the
near-rigid precession requirement.



\section{Summary}\label{sec:sum}

We have reported the first XRISM observation of the dwarf galaxy NGC~4395,
which hosts a low-mass AGN, complemented by a simultaneous NuSTAR observation.
Using XRISM/Resolve, we investigated the Fe~K band around 6~keV after robustly
determining the continuum shape from a joint fit to the Resolve and NuSTAR
spectra, excluding the 5.5--7.5~keV range. Our main results are summarized as
follows.

\begin{itemize}
\item The time-averaged Resolve spectrum shows an unresolved neutral Fe~K$\alpha$
core with a velocity width of $\lesssim110$~km~s$^{-1}$, accompanied by a
redward wing (Figure~\ref{fig:first_res_spec}). The redward component is well
reproduced by an additional relativistically broadened Fe~K component. In the
\texttt{MYTorus}-based fits, representative choices of
$R_{\rm out}=2\times10^2\,R_{\rm g}$ and $10^3\,R_{\rm g}$ give consistent
parameters, with $R_{\rm in}\sim70\,R_{\rm g}$.

\item
By splitting the Resolve exposure into two equal time intervals, we found that
the broadened Fe~K component varied significantly within the $\sim400$~ks
observation (Section~\ref{sec:twobin}, Table~\ref{tab:time_res_fitted_pars}, and Figure~\ref{fig:spec_o2}).

\item
With five finer time bins, fits in which both $R_{\rm in}$ and $\theta_{\rm inc}$ were allowed to vary show their significant changes (Section~\ref{sec:finebin} and Figure~\ref{fig:five_bin_par_changes}).

\item The radial inward evolution of the line-emitting region is qualitatively consistent with the
contemporaneous brightening of the power-law, or Comptonized, continuum
emission, plausibly due to an increased supply of disk seed photons as the
inner disk moves inward.

\item 
We also found a possible periodic trend in the apparent inclination angle, with
a period of $213\pm9$~ks. If this modulation is interpreted as LT precession
of a fiducial compact region with $R_{\rm in}\simeq100\,R_{\rm g}$ and
$R_{\rm out}=130\,R_{\rm g}$, the observed period favors the lower end
of the BH-mass estimates and requires a high prograde spin of $a\sim0.9$.
Even in an extremely compact configuration, a prograde spin of
$a\gtrsim0.6$ is still required.
\end{itemize}

If confirmed, the quasi-periodic modulation of the apparent inclination of the
broadened Fe~K component would provide a direct X-ray spectroscopic signature
of LT precession in an AGN. Such a signature is more readily detectable in low-mass systems such as NGC~4395 than in more massive Seyfert galaxies, because their shorter
relativistic timescales make geometric changes in the inner accretion flow
observable within a single observation.
Such measurements would open a new way to probe frame dragging and the
three-dimensional inner accretion geometry, offering spin constraints
complementary to those from broadband reflection fitting.

\begin{acknowledgments}

We thank the anonymous referee for constructive comments and suggestions that improved the clarity and robustness of this work.
TK is grateful to Claudio Ricci for fruitful discussion. 
Part of this work was supported by JSPS KAKENHI Grant Number 23K13153/24K00673 (TK), 23K13154 (SY),   23K20239/24K00672/25H00660 (HN), 22K18277/23KF0254/26H00604 (YI), 
24K17104 (SO), 21K13958 (MM), and Yamada Science Foundation (MM).

This research has made use of data and/or software provided by the High Energy Astrophysics Science Archive Research Center (HEASARC), which is a service of the Astrophysics Science Division at NASA/GSFC.
This research has made use of data from the NuSTAR mission, a project led by the California Institute of Technology, managed by the Jet Propulsion Laboratory, and funded by the National Aeronautics and Space Administration. Data analysis was performed using the NuSTAR Data Analysis Software (NuSTARDAS), jointly developed by the ASI Science Data Center (SSDC, Italy) and the California Institute of Technology (USA).
\end{acknowledgments}

\appendix

\section{Dependence of $C$-statistic Diagnostics on Spectral Binning and Photon Statistics}
\label{app:binning}

In this appendix, we examine how the $C$-statistic diagnostics used in this
paper, $C_{\rm obs}/C_{\rm exp}$ and
$(C_{\rm obs}-C_{\rm exp})/\sqrt{C_v}$, depend on the adopted spectral grouping
and on the number of source counts.
This examination is useful because the expected value of the $C$ statistic depends on the model-predicted counts in each spectral bin \citep{Kaa17}, and therefore can change with both the binning scheme and the photon statistics of the data. 
In addition, the minimized $C$ statistic 
after spectral fitting can be shifted from the value expected before
optimization. We therefore performed Monte Carlo simulations to quantify the
behavior of $C_{\rm obs}/C_{\rm exp}$ and
$(C_{\rm obs}-C_{\rm exp})/\sqrt{C_{\rm v}}$ under several grouping
prescriptions and source-count levels.

We considered three classes of grouping. 
First, we used optimal 
binning with \texttt{optmin}$=1$, 10, and 20. 
Second, we used simple minimum-count groupings, requiring at least $g_{\rm min}=1$, 10, and 20. Third, to
consider brighter spectra or higher counts per bin, 
we repeated the $g_{\rm min} =1$ simulations after
increasing the source normalization of the parent model by factors of 10 and
50. For each setup, we generated 1000 fake spectra from the same reference
model, regrouped each realization with the corresponding prescription, and
fitted it with the same model.
The reference model was the time-averaged best-fit diskline model M4a in
Table~\ref{tab:model_ladder}, composed of the baseline continuum, 
\texttt{zbfeklor}+\texttt{zbfekblor} for the narrow neutral Fe~K$\alpha$ and K$\beta$ lines, and the \texttt{diskline} component for the broadened Fe~K$\alpha$ emission with $R_{\rm out}=2\times10^2\,R_{\rm g}$. We adopted this
computationally reasonable model instead of the final
\texttt{rdblur*MYTorusL}-based model.

\begin{figure*}
  \centering 
    \includegraphics[width=0.48\textwidth]{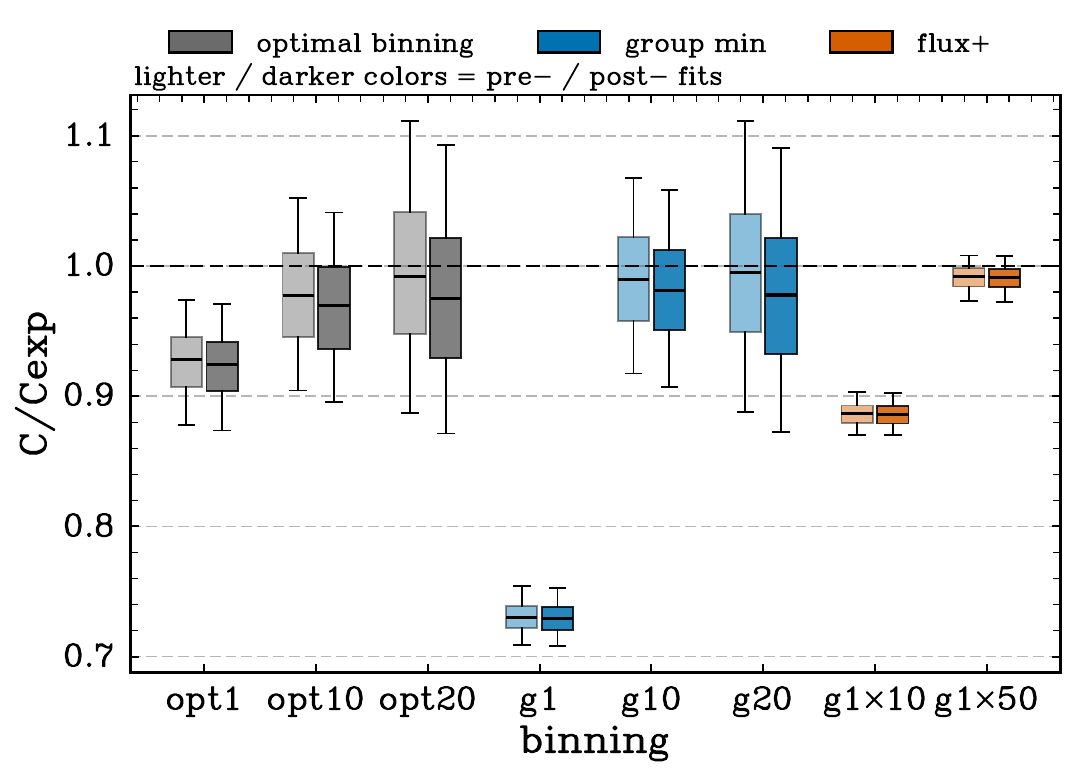}
    \includegraphics[width=0.48\textwidth]{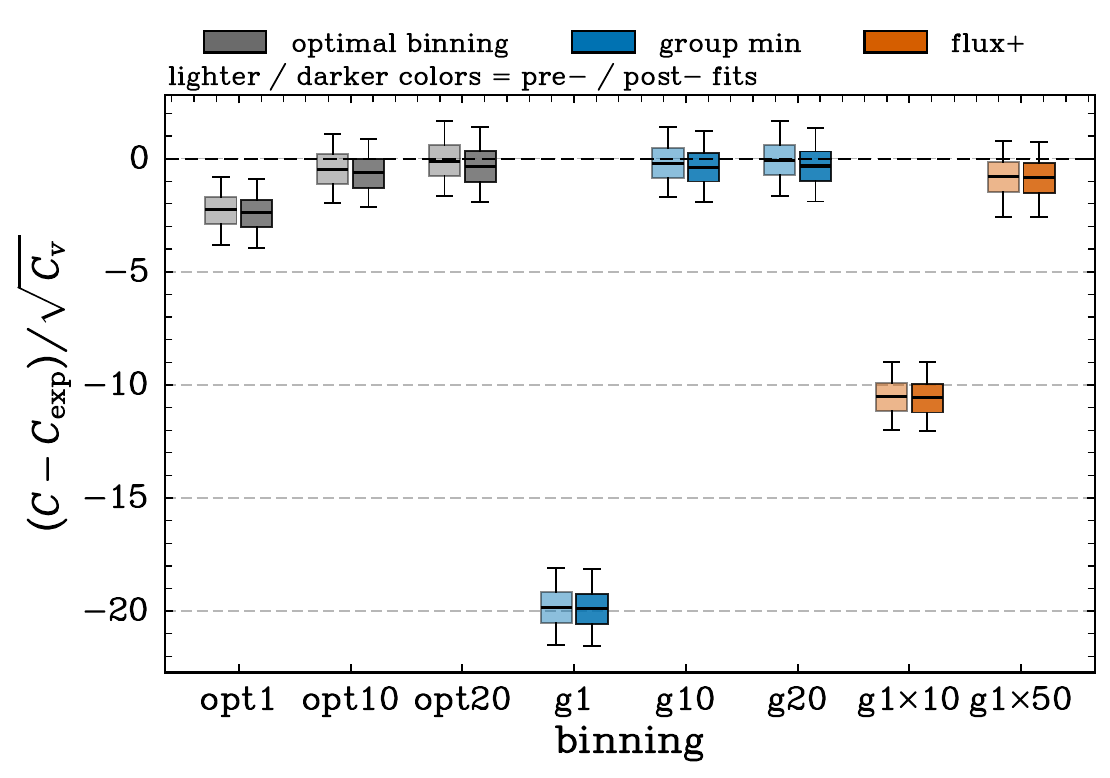}
    \includegraphics[width=0.48\textwidth]{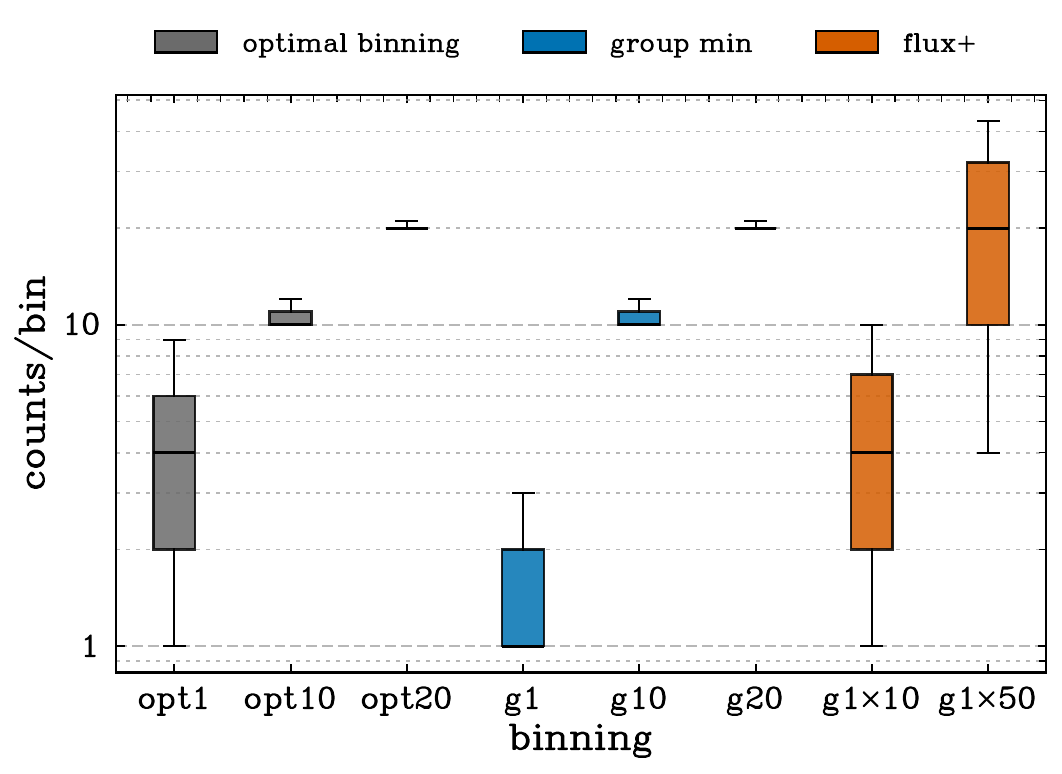} 
      \vspace{-0.3cm}  
  \caption{
Results of the Monte Carlo tests for different spectral grouping prescriptions.
Top left: distributions of
$C_{\rm prefit}/C_{\rm exp,prefit}$ and
$C_{\rm postfit}/C_{\rm exp,postfit}$.
Top right: distributions of the normalized statistic,
$(C-C_{\rm exp})/\sqrt{C_{\rm v}}$.
Bottom: distributions of the number of counts per spectral bin.
Gray, blue, and orange boxes denote optimal binning, fixed minimum-count
grouping, and flux-amplified cases, respectively.
Along the horizontal axis, \texttt{opt1}--\texttt{opt20} represent optimal
binning with \texttt{optmin}$=1$, 10, and 20;
\texttt{g1}--\texttt{g20} represent fixed minimum-count grouping with
$g_{\rm min}=1$, 10, and 20; and
\texttt{g1}$\times$10 and \texttt{g1}$\times$50 represent the
flux-amplified runs with the source normalization increased by factors of
10 and 50.
For the top left and top right panels, the lighter/thinner boxes show the
pre-fit results, while the darker/thicker boxes show the post-fit results.
Boxes indicate the interquartile range, the central black line marks the
median, and the whiskers denote the 5th and 95th percentiles.
}\label{fig:c_vs_binning}  
\end{figure*}

For each simulated spectrum, we computed the $C$ statistic in two ways. The
first is the pre-fit value, $C_{\rm prefit}$, evaluated using the input model
before refitting the simulated spectrum. The second is the post-fit value,
$C_{\rm postfit}$, obtained after refitting the simulated spectrum. We then
computed the corresponding expected $C$ statistic ($C_{\rm exp}$) from the model, and its
variance, $C_{\rm v}$, following \citet{Kaa17}. 
The comparison between
the pre-fit and post-fit values allows us to examine changes in the $C$ values 
due to likelihood minimization.

Figure~\ref{fig:c_vs_binning} shows the resulting distributions of
$C_{\rm prefit}/C_{\rm exp,prefit}$ and
$C_{\rm postfit}/C_{\rm exp,postfit}$, together with the distributions of
$(C-C_{\rm exp})/\sqrt{C_{\rm v}}$ and the number of counts per spectral bin.
Comparing the pre-fit and post-fit results, the post-fit distributions are
shifted toward slightly smaller $C/C_{\rm exp}$ values, as expected because
the likelihood optimization minimizes the $C$ statistic for each simulated
spectrum. However, this shift is modest compared with the systematic
differences caused by the adopted grouping prescription.

The dominant trend is instead the dependence on the minimum number of counts
allowed in each spectral bin. For both the optimal-binning and simple
minimum-count groupings, smaller values of \texttt{optmin} or \texttt{gmin}
lead to smaller $C/C_{\rm exp}$ and more negative
$(C-C_{\rm exp})/\sqrt{C_{\rm v}}$. 
This trend indicates that spectra grouped
more finely, and hence containing lower-count bins, tend to yield $C$-statistic
values below the expected value. The effect is particularly strong for the
$g_{\rm min} = 1$ case. As shown by the counts-per-bin distribution in the bottom 
panel, this grouping produces spectra in which most bins contain only one
count, and the typical counts per bin are much smaller than in the other
grouping prescriptions. In this regime, $C/C_{\rm exp}$ can become
substantially smaller than unity even though the simulated spectra are
generated from the same model used in the fit.

The flux-amplified simulations clarify that this behavior is mainly driven by
the low counts per bin, rather than by the use of $g_{\rm min}=1$ itself.
When the source normalization is increased by factors of 10 and 50, the
$g_{\rm min}=1$ spectra contain more bins with more than one count already at
the native-bin level. Correspondingly, the distributions of $C/C_{\rm exp}$
move closer to unity and
$(C-C_{\rm exp})/\sqrt{C_{\rm v}}$ becomes closer to zero. 
Thus, a small value of $C/C_{\rm exp}$, or equivalently a small value of $C/{\rm d.o.f.}$, can naturally occur when the spectrum is dominated by one-count bins, whereas the diagnostic approaches the expected behavior as the counts per bin increase.





%
\facilities{XRISM \citep{Tashiro2025PASJ...77S...1T}, NuSTAR \citep{Har13}}


\software{Astropy \citep{astropy:2013, astropy:2018, astropy:2022}} 


\bibliography{ref}{}

@INBOOK{Arn96, 
       author = {{Arnaud}, K.~A.},
        title = "{XSPEC: The First Ten Years}",
    booktitle = {Astronomical Data Analysis Software and Systems V, A.S.P. Conference Series, Vol. 101, 1996, George H. Jacoby and Jeannette Barnes, eds., p. 17.},
         year = "1996",
       editor = {{Jacoby}, George H. and {Barnes}, Jeannette},
       volume = {101},
       series = {Astronomical Society of the Pacific Conference Series},
        pages = {17},
       adsurl = {https://ui.adsabs.harvard.edu/abs/1996ASPC..101...17A}
}

@ARTICLE{Cas79, 
       author = {{Cash}, W.},
        title = "{Parameter estimation in astronomy through application of the likelihood ratio.}",
      journal = {\apj},
         year = "1979",
        month = "Mar",
       volume = {228},
        pages = {939-947},
          doi = {10.1086/156922},
       adsurl = {https://ui.adsabs.harvard.edu/abs/1979ApJ...228..939C}
}

@ARTICLE{Har13,
       author = {{Harrison}, Fiona A. and {Craig}, William W. and {Christensen}, Finn E. and {Hailey}, Charles J. and {Zhang}, William W. and {Boggs}, Steven E. and {Stern}, Daniel and {Cook}, W. Rick and {Forster}, Karl and {Giommi}, Paolo and {Grefenstette}, Brian W. and {Kim}, Yunjin and {Kitaguchi}, Takao and {Koglin}, Jason E. and {Madsen}, Kristin K. and {Mao}, Peter H. and {Miyasaka}, Hiromasa and {Mori}, Kaya and {Perri}, Matteo and {Pivovaroff}, Michael J. and {Puccetti}, Simonetta and {Rana}, Vikram R. and {Westergaard}, Niels J. and {Willis}, Jason and {Zoglauer}, Andreas and {An}, Hongjun and {Bachetti}, Matteo and {Barri{\`e}re}, Nicolas M. and {Bellm}, Eric C. and {Bhalerao}, Varun and {Brejnholt}, Nicolai F. and {Fuerst}, Felix and {Liebe}, Carl C. and {Markwardt}, Craig B. and {Nynka}, Melania and {Vogel}, Julia K. and {Walton}, Dominic J. and {Wik}, Daniel R. and {Alexander}, David M. and {Cominsky}, Lynn R. and {Hornschemeier}, Ann E. and {Hornstrup}, Allan and {Kaspi}, Victoria M. and {Madejski}, Greg M. and {Matt}, Giorgio and {Molendi}, Silvano and {Smith}, David M. and {Tomsick}, John A. and {Ajello}, Marco and {Ballantyne}, David R. and {Balokovi{\'c}}, Mislav and {Barret}, Didier and {Bauer}, Franz E. and {Blandford}, Roger D. and {Brandt}, W. Niel and {Brenneman}, Laura W. and {Chiang}, James and {Chakrabarty}, Deepto and {Chenevez}, Jerome and {Comastri}, Andrea and {Dufour}, Francois and {Elvis}, Martin and {Fabian}, Andrew C. and {Farrah}, Duncan and {Fryer}, Chris L. and {Gotthelf}, Eric V. and {Grindlay}, Jonathan E. and {Helfand}, David J. and {Krivonos}, Roman and {Meier}, David L. and {Miller}, Jon M. and {Natalucci}, Lorenzo and {Ogle}, Patrick and {Ofek}, Eran O. and {Ptak}, Andrew and {Reynolds}, Stephen P. and {Rigby}, Jane R. and {Tagliaferri}, Gianpiero and {Thorsett}, Stephen E. and {Treister}, Ezequiel and {Urry}, C. Megan},
        title = "{The Nuclear Spectroscopic Telescope Array (NuSTAR) High-energy X-Ray Mission}",
      journal = {\apj},
         year = 2013,
        month = jun,
       volume = {770},
       number = {2},
          eid = {103},
        pages = {103},
          doi = {10.1088/0004-637X/770/2/103},
archivePrefix = {arXiv},
       eprint = {1301.7307},
 primaryClass = {astro-ph.IM},
       adsurl = {https://ui.adsabs.harvard.edu/abs/2013ApJ...770..103H}
}

@ARTICLE{Kaa17, 
       author = {{Kaastra}, J.~S.},
        title = "{On the use of C-stat in testing models for X-ray spectra}",
      journal = {\aap},
         year = "2017",
        month = "Sep",
       volume = {605},
          eid = {A51},
        pages = {A51},
          doi = {10.1051/0004-6361/201629319},
archivePrefix = {arXiv},
       eprint = {1707.09202},
 primaryClass = {astro-ph.HE},
       adsurl = {https://ui.adsabs.harvard.edu/abs/2017A&A...605A..51K}
}

@ARTICLE{Mur09, 
       author = {{Murphy}, Kendrah D. and {Yaqoob}, Tahir},
        title = "{An X-ray spectral model for Compton-thick toroidal reprocessors}",
      journal = {\mnras},
         year = "2009",
        month = "Aug",
       volume = {397},
       number = {3},
        pages = {1549-1562},
          doi = {10.1111/j.1365-2966.2009.15025.x},
archivePrefix = {arXiv},
       eprint = {0905.3188},
 primaryClass = {astro-ph.HE},
       adsurl = {https://ui.adsabs.harvard.edu/abs/2009MNRAS.397.1549M}
}

@ARTICLE{Ric17bass,
       author = {{Ricci}, C. and {Trakhtenbrot}, B. and {Koss}, M.~J. and {Ueda}, Y. and {Del Vecchio}, I. and {Treister}, E. and {Schawinski}, K. and {Paltani}, S. and {Oh}, K. and {Lamperti}, I. and {Berney}, S. and {Gandhi}, P. and {Ichikawa}, K. and {Bauer}, F.~E. and {Ho}, L.~C. and {Asmus}, D. and {Beckmann}, V. and {Soldi}, S. and {Balokovi{\'c}}, M. and {Gehrels}, N. and {Markwardt}, C.~B.},
        title = "{BAT AGN Spectroscopic Survey. V. X-Ray Properties of the Swift/BAT 70-month AGN Catalog}",
      journal = {\apjs},
         year = 2017,
        month = dec,
       volume = {233},
       number = {2},
          eid = {17},
        pages = {17},
          doi = {10.3847/1538-4365/aa96ad},
archivePrefix = {arXiv},
       eprint = {1709.03989},
 primaryClass = {astro-ph.HE},
       adsurl = {https://ui.adsabs.harvard.edu/abs/2017ApJS..233...17R}
}

@ARTICLE{Tan19,
       author = {{Tanimoto}, Atsushi and {Ueda}, Yoshihiro and {Odaka}, Hirokazu and {Kawaguchi}, Toshihiro and {Fukazawa}, Yasushi and {Kawamuro}, Taiki},
        title = "{XCLUMPY: X-Ray Spectral Model from Clumpy Torus and Its Application to the Circinus Galaxy}",
      journal = {\apj},
         year = 2019,
        month = jun,
       volume = {877},
       number = {2},
          eid = {95},
        pages = {95},
          doi = {10.3847/1538-4357/ab1b20},
archivePrefix = {arXiv},
       eprint = {1904.08945},
 primaryClass = {astro-ph.HE},
       adsurl = {https://ui.adsabs.harvard.edu/abs/2019ApJ...877...95T}
}

@ARTICLE{Hol_1997PhRvA..56.4554H,
       author = {{H{\"o}lzer}, G. and {Fritsch}, M. and {Deutsch}, M. and {H{\"a}rtwig}, J. and {F{\"o}rster}, E.},
        title = "{K{\ensuremath{\alpha}}$_{1,2}$ and K{\ensuremath{\beta}}$_{1,3}$ x-ray emission lines of the 3d transition metals}",
      journal = {\pra},
         year = 1997,
        month = dec,
       volume = {56},
       number = {6},
        pages = {4554-4568},
          doi = {10.1103/PhysRevA.56.4554},
       adsurl = {https://ui.adsabs.harvard.edu/abs/1997PhRvA..56.4554H}
}

@article{astropy:2013,
Adsurl = {http://adsabs.harvard.edu/abs/2013A%26A...558A..33A},
Archiveprefix = {arXiv},
Author = {{Astropy Collaboration} and {Robitaille}, T.~P. and {Tollerud}, E.~J. and {Greenfield}, P. and {Droettboom}, M. and {Bray}, E. and {Aldcroft}, T. and {Davis}, M. and {Ginsburg}, A. and {Price-Whelan}, A.~M. and {Kerzendorf}, W.~E. and {Conley}, A. and {Crighton}, N. and {Barbary}, K. and {Muna}, D. and {Ferguson}, H. and {Grollier}, F. and {Parikh}, M.~M. and {Nair}, P.~H. and {Unther}, H.~M. and {Deil}, C. and {Woillez}, J. and {Conseil}, S. and {Kramer}, R. and {Turner}, J.~E.~H. and {Singer}, L. and {Fox}, R. and {Weaver}, B.~A. and {Zabalza}, V. and {Edwards}, Z.~I. and {Azalee Bostroem}, K. and {Burke}, D.~J. and {Casey}, A.~R. and {Crawford}, S.~M. and {Dencheva}, N. and {Ely}, J. and {Jenness}, T. and {Labrie}, K. and {Lim}, P.~L. and {Pierfederici}, F. and {Pontzen}, A. and {Ptak}, A. and {Refsdal}, B. and {Servillat}, M. and {Streicher}, O.},
Doi = {10.1051/0004-6361/201322068},
Eid = {A33},
Eprint = {1307.6212},
Journal = {\aap},
Month = oct,
Pages = {A33},
Primaryclass = {astro-ph.IM},
Title = {{Astropy: A community Python package for astronomy}},
Volume = 558,
Year = 2013}

@ARTICLE{astropy:2018,
       author = {{Astropy Collaboration} and {Price-Whelan}, A.~M. and
         {Sip{\H{o}}cz}, B.~M. and {G{\"u}nther}, H.~M. and {Lim}, P.~L. and
         {Crawford}, S.~M. and {Conseil}, S. and {Shupe}, D.~L. and
         {Craig}, M.~W. and {Dencheva}, N. and {Ginsburg}, A. and {Vand
        erPlas}, J.~T. and {Bradley}, L.~D. and {P{\'e}rez-Su{\'a}rez}, D. and
         {de Val-Borro}, M. and {Aldcroft}, T.~L. and {Cruz}, K.~L. and
         {Robitaille}, T.~P. and {Tollerud}, E.~J. and {Ardelean}, C. and
         {Babej}, T. and {Bach}, Y.~P. and {Bachetti}, M. and {Bakanov}, A.~V. and
         {Bamford}, S.~P. and {Barentsen}, G. and {Barmby}, P. and
         {Baumbach}, A. and {Berry}, K.~L. and {Biscani}, F. and {Boquien}, M. and
         {Bostroem}, K.~A. and {Bouma}, L.~G. and {Brammer}, G.~B. and
         {Bray}, E.~M. and {Breytenbach}, H. and {Buddelmeijer}, H. and
         {Burke}, D.~J. and {Calderone}, G. and {Cano Rodr{\'\i}guez}, J.~L. and
         {Cara}, M. and {Cardoso}, J.~V.~M. and {Cheedella}, S. and {Copin}, Y. and
         {Corrales}, L. and {Crichton}, D. and {D'Avella}, D. and {Deil}, C. and
         {Depagne}, {\'E}. and {Dietrich}, J.~P. and {Donath}, A. and
         {Droettboom}, M. and {Earl}, N. and {Erben}, T. and {Fabbro}, S. and
         {Ferreira}, L.~A. and {Finethy}, T. and {Fox}, R.~T. and
         {Garrison}, L.~H. and {Gibbons}, S.~L.~J. and {Goldstein}, D.~A. and
         {Gommers}, R. and {Greco}, J.~P. and {Greenfield}, P. and
         {Groener}, A.~M. and {Grollier}, F. and {Hagen}, A. and {Hirst}, P. and
         {Homeier}, D. and {Horton}, A.~J. and {Hosseinzadeh}, G. and {Hu}, L. and
         {Hunkeler}, J.~S. and {Ivezi{\'c}}, {\v{Z}}. and {Jain}, A. and
         {Jenness}, T. and {Kanarek}, G. and {Kendrew}, S. and {Kern}, N.~S. and
         {Kerzendorf}, W.~E. and {Khvalko}, A. and {King}, J. and {Kirkby}, D. and
         {Kulkarni}, A.~M. and {Kumar}, A. and {Lee}, A. and {Lenz}, D. and
         {Littlefair}, S.~P. and {Ma}, Z. and {Macleod}, D.~M. and
         {Mastropietro}, M. and {McCully}, C. and {Montagnac}, S. and
         {Morris}, B.~M. and {Mueller}, M. and {Mumford}, S.~J. and {Muna}, D. and
         {Murphy}, N.~A. and {Nelson}, S. and {Nguyen}, G.~H. and
         {Ninan}, J.~P. and {N{\"o}the}, M. and {Ogaz}, S. and {Oh}, S. and
         {Parejko}, J.~K. and {Parley}, N. and {Pascual}, S. and {Patil}, R. and
         {Patil}, A.~A. and {Plunkett}, A.~L. and {Prochaska}, J.~X. and
         {Rastogi}, T. and {Reddy Janga}, V. and {Sabater}, J. and
         {Sakurikar}, P. and {Seifert}, M. and {Sherbert}, L.~E. and
         {Sherwood-Taylor}, H. and {Shih}, A.~Y. and {Sick}, J. and
         {Silbiger}, M.~T. and {Singanamalla}, S. and {Singer}, L.~P. and
         {Sladen}, P.~H. and {Sooley}, K.~A. and {Sornarajah}, S. and
         {Streicher}, O. and {Teuben}, P. and {Thomas}, S.~W. and
         {Tremblay}, G.~R. and {Turner}, J.~E.~H. and {Terr{\'o}n}, V. and
         {van Kerkwijk}, M.~H. and {de la Vega}, A. and {Watkins}, L.~L. and
         {Weaver}, B.~A. and {Whitmore}, J.~B. and {Woillez}, J. and
         {Zabalza}, V. and {Astropy Contributors}},
        title = "{The Astropy Project: Building an Open-science Project and Status of the v2.0 Core Package}",
      journal = {\aj},
         year = 2018,
        month = sep,
       volume = {156},
       number = {3},
          eid = {123},
        pages = {123},
          doi = {10.3847/1538-3881/aabc4f},
archivePrefix = {arXiv},
       eprint = {1801.02634},
 primaryClass = {astro-ph.IM},
       adsurl = {https://ui.adsabs.harvard.edu/abs/2018AJ....156..123A}
}

@ARTICLE{astropy:2022,
       author = {{Astropy Collaboration} and {Price-Whelan}, Adrian M. and {Lim}, Pey Lian and {Earl}, Nicholas and {Starkman}, Nathaniel and {Bradley}, Larry and {Shupe}, David L. and {Patil}, Aarya A. and {Corrales}, Lia and {Brasseur}, C.~E. and {N{"o}the}, Maximilian and {Donath}, Axel and {Tollerud}, Erik and {Morris}, Brett M. and {Ginsburg}, Adam and {Vaher}, Eero and {Weaver}, Benjamin A. and {Tocknell}, James and {Jamieson}, William and {van Kerkwijk}, Marten H. and {Robitaille}, Thomas P. and {Merry}, Bruce and {Bachetti}, Matteo and {G{"u}nther}, H. Moritz and {Aldcroft}, Thomas L. and {Alvarado-Montes}, Jaime A. and {Archibald}, Anne M. and {B{'o}di}, Attila and {Bapat}, Shreyas and {Barentsen}, Geert and {Baz{'a}n}, Juanjo and {Biswas}, Manish and {Boquien}, M{'e}d{'e}ric and {Burke}, D.~J. and {Cara}, Daria and {Cara}, Mihai and {Conroy}, Kyle E. and {Conseil}, Simon and {Craig}, Matthew W. and {Cross}, Robert M. and {Cruz}, Kelle L. and {D'Eugenio}, Francesco and {Dencheva}, Nadia and {Devillepoix}, Hadrien A.~R. and {Dietrich}, J{"o}rg P. and {Eigenbrot}, Arthur Davis and {Erben}, Thomas and {Ferreira}, Leonardo and {Foreman-Mackey}, Daniel and {Fox}, Ryan and {Freij}, Nabil and {Garg}, Suyog and {Geda}, Robel and {Glattly}, Lauren and {Gondhalekar}, Yash and {Gordon}, Karl D. and {Grant}, David and {Greenfield}, Perry and {Groener}, Austen M. and {Guest}, Steve and {Gurovich}, Sebastian and {Handberg}, Rasmus and {Hart}, Akeem and {Hatfield-Dodds}, Zac and {Homeier}, Derek and {Hosseinzadeh}, Griffin and {Jenness}, Tim and {Jones}, Craig K. and {Joseph}, Prajwel and {Kalmbach}, J. Bryce and {Karamehmetoglu}, Emir and {Ka{l}uszy{'n}ski}, Miko{l}aj and {Kelley}, Michael S.~P. and {Kern}, Nicholas and {Kerzendorf}, Wolfgang E. and {Koch}, Eric W. and {Kulumani}, Shankar and {Lee}, Antony and {Ly}, Chun and {Ma}, Zhiyuan and {MacBride}, Conor and {Maljaars}, Jakob M. and {Muna}, Demitri and {Murphy}, N.~A. and {Norman}, Henrik and {O'Steen}, Richard and {Oman}, Kyle A. and {Pacifici}, Camilla and {Pascual}, Sergio and {Pascual-Granado}, J. and {Patil}, Rohit R. and {Perren}, Gabriel I. and {Pickering}, Timothy E. and {Rastogi}, Tanuj and {Roulston}, Benjamin R. and {Ryan}, Daniel F. and {Rykoff}, Eli S. and {Sabater}, Jose and {Sakurikar}, Parikshit and {Salgado}, Jes{'u}s and {Sanghi}, Aniket and {Saunders}, Nicholas and {Savchenko}, Volodymyr and {Schwardt}, Ludwig and {Seifert-Eckert}, Michael and {Shih}, Albert Y. and {Jain}, Anany Shrey and {Shukla}, Gyanendra and {Sick}, Jonathan and {Simpson}, Chris and {Singanamalla}, Sudheesh and {Singer}, Leo P. and {Singhal}, Jaladh and {Sinha}, Manodeep and {Sip{H{o}}cz}, Brigitta M. and {Spitler}, Lee R. and {Stansby}, David and {Streicher}, Ole and {{{S}}umak}, Jani and {Swinbank}, John D. and {Taranu}, Dan S. and {Tewary}, Nikita and {Tremblay}, Grant R. and {Val-Borro}, Miguel de and {Van Kooten}, Samuel J. and {Vasovi{'c}}, Zlatan and {Verma}, Shresth and {de Miranda Cardoso}, Jos{'e} Vin{'i}cius and {Williams}, Peter K.~G. and {Wilson}, Tom J. and {Winkel}, Benjamin and {Wood-Vasey}, W.~M. and {Xue}, Rui and {Yoachim}, Peter and {Zhang}, Chen and {Zonca}, Andrea and {Astropy Project Contributors}},
        title = "{The Astropy Project: Sustaining and Growing a Community-oriented Open-source Project and the Latest Major Release (v5.0) of the Core Package}",
      journal = {\apj},
         year = 2022,
        month = aug,
       volume = {935},
       number = {2},
          eid = {167},
        pages = {167},
          doi = {10.3847/1538-4357/ac7c74},
archivePrefix = {arXiv},
       eprint = {2206.14220},
 primaryClass = {astro-ph.IM},
       adsurl = {https://ui.adsabs.harvard.edu/abs/2022ApJ...935..167A}
}

@ARTICLE{McHardy2023MNRAS.519.3366M,
       author = {{McHardy}, I.~M. and {Beard}, M. and {Breedt}, E. and {Knapen}, J.~H. and {Vincentelli}, F.~M. and {Veresvarska}, M. and {Dhillon}, V.~S. and {Marsh}, T.~R. and {Littlefair}, S.~P. and {Horne}, K. and {Glew}, R. and {Goad}, M.~R. and {Kammoun}, E. and {Emmanoulopoulos}, D.},
        title = "{First detection of the outer edge of an AGN accretion disc: very fast multiband optical variability of NGC 4395 with GTC/HiPERCAM and LT/IO:O}",
      journal = {\mnras},
         year = 2023,
        month = mar,
       volume = {519},
       number = {3},
        pages = {3366-3382},
          doi = {10.1093/mnras/stac3651},
archivePrefix = {arXiv},
       eprint = {2212.08015},
 primaryClass = {astro-ph.GA},
       adsurl = {https://ui.adsabs.harvard.edu/abs/2023MNRAS.519.3366M}
}

@ARTICLE{Franchini2016MNRAS.455.1946F,
       author = {{Franchini}, Alessia and {Lodato}, Giuseppe and {Facchini}, Stefano},
        title = "{Lense-Thirring precession around supermassive black holes during tidal disruption events}",
      journal = {\mnras},
         year = 2016,
        month = jan,
       volume = {455},
       number = {2},
        pages = {1946-1956},
          doi = {10.1093/mnras/stv2417},
archivePrefix = {arXiv},
       eprint = {1510.04879},
 primaryClass = {astro-ph.HE},
       adsurl = {https://ui.adsabs.harvard.edu/abs/2016MNRAS.455.1946F}
}

@ARTICLE{Woo2019NatAs...3..755W,
       author = {{Woo}, Jong-Hak and {Cho}, Hojin and {Gallo}, Elena and {Hodges-Kluck}, Edmund and {Le}, Huynh Anh N. and {Shin}, Jaejin and {Son}, Donghoon and {Horst}, John C.},
        title = "{A 10,000-solar-mass black hole in the nucleus of a bulgeless dwarf galaxy}",
      journal = {Nature Astronomy},
         year = 2019,
        month = jun,
       volume = {3},
        pages = {755-759},
          doi = {10.1038/s41550-019-0790-3},
archivePrefix = {arXiv},
       eprint = {1905.00145},
 primaryClass = {astro-ph.GA},
       adsurl = {https://ui.adsabs.harvard.edu/abs/2019NatAs...3..755W}
}

@ARTICLE{Peterson2005ApJ...632..799P,
       author = {{Peterson}, Bradley M. and {Bentz}, Misty C. and {Desroches}, Louis-Benoit and {Filippenko}, Alexei V. and {Ho}, Luis C. and {Kaspi}, Shai and {Laor}, Ari and {Maoz}, Dan and {Moran}, Edward C. and {Pogge}, Richard W. and {Quillen}, Alice C.},
        title = "{Multiwavelength Monitoring of the Dwarf Seyfert 1 Galaxy NGC 4395. I. A Reverberation-based Measurement of the Black Hole Mass}",
      journal = {\apj},
         year = 2005,
        month = oct,
       volume = {632},
       number = {2},
        pages = {799-808},
          doi = {10.1086/444494},
archivePrefix = {arXiv},
       eprint = {astro-ph/0506665},
 primaryClass = {astro-ph},
       adsurl = {https://ui.adsabs.harvard.edu/abs/2005ApJ...632..799P}
}

@ARTICLE{Iwasawa2010A&A...514A..58I,
       author = {{Iwasawa}, K. and {Tanaka}, Y. and {Gallo}, L.~C.},
        title = "{The Suzaku broadband X-ray spectrum of the dwarf Seyfert galaxy NGC 4395}",
      journal = {\aap},
         year = 2010,
        month = may,
       volume = {514},
          eid = {A58},
        pages = {A58},
          doi = {10.1051/0004-6361/200912431},
archivePrefix = {arXiv},
       eprint = {1002.2062},
 primaryClass = {astro-ph.CO},
       adsurl = {https://ui.adsabs.harvard.edu/abs/2010A&A...514A..58I}
}

@ARTICLE{Vaughan2005MNRAS.356..524V,
       author = {{Vaughan}, S. and {Iwasawa}, K. and {Fabian}, A.~C. and {Hayashida}, K.},
        title = "{The exceptional X-ray variability of the dwarf Seyfert nucleus NGC 4395}",
      journal = {\mnras},
         year = 2005,
        month = jan,
       volume = {356},
       number = {2},
        pages = {524-530},
          doi = {10.1111/j.1365-2966.2004.08463.x},
archivePrefix = {arXiv},
       eprint = {astro-ph/0410261},
 primaryClass = {astro-ph},
       adsurl = {https://ui.adsabs.harvard.edu/abs/2005MNRAS.356..524V}
}

@ARTICLE{George1991MNRAS.249..352G,
       author = {{George}, I.~M. and {Fabian}, A.~C.},
        title = "{X-ray reflection from cold matter in Active Galactic Nuclei and X-ray binaries.}",
      journal = {\mnras},
         year = 1991,
        month = mar,
       volume = {249},
        pages = {352},
          doi = {10.1093/mnras/249.2.352},
       adsurl = {https://ui.adsabs.harvard.edu/abs/1991MNRAS.249..352G}
}

@ARTICLE{Xrism2024ApJ...973L..25X,
       author = {{XRISM Collaboration} and {Audard}, Marc and {Awaki}, Hisamitsu and {Ballhausen}, Ralf and {Bamba}, Aya and {Behar}, Ehud and {Boissay-Malaquin}, Rozenn and {Brenneman}, Laura and {Brown}, Gregory V. and {Corrales}, Lia and {Costantini}, Elisa and {Cumbee}, Renata and {Diaz Trigo}, Maria and {Done}, Chris and {Dotani}, Tadayasu and {Ebisawa}, Ken and {Eckart}, Megan E. and {Eckert}, Dominique and {Enoto}, Teruaki and {Eguchi}, Satoshi and {Ezoe}, Yuichiro and {Foster}, Adam and {Fujimoto}, Ryuichi and {Fujita}, Yutaka and {Fukazawa}, Yasushi and {Fukushima}, Kotaro and {Furuzawa}, Akihiro and {Gallo}, Luigi and {Garc{\'\i}a}, Javier A. and {Gu}, Liyi and {Guainazzi}, Matteo and {Hagino}, Kouichi and {Hamaguchi}, Kenji and {Hatsukade}, Isamu and {Hayashi}, Katsuhiro and {Hayashi}, Takayuki and {Hell}, Natalie and {Hodges-Kluck}, Edmund and {Hornschemeier}, Ann and {Ichinohe}, Yuto and {Ishida}, Manabu and {Ishikawa}, Kumi and {Ishisaki}, Yoshitaka and {Kaastra}, Jelle and {Kallman}, Timothy and {Kara}, Erin and {Katsuda}, Satoru and {Kanemaru}, Yoshiaki and {Kelley}, Richard and {Kilbourne}, Caroline and {Kitamoto}, Shunji and {Kobayashi}, Shogo and {Kohmura}, Takayoshi and {Kubota}, Aya and {Leutenegger}, Maurice and {Loewenstein}, Michael and {Maeda}, Yoshitomo and {Markevitch}, Maxim and {Matsumoto}, Hironori and {Matsushita}, Kyoko and {McCammon}, Dan and {McNamara}, Brian and {Mernier}, Fran{\c{c}}ois and {Miller}, Eric D. and {Miller}, Jon M. and {Mitsuishi}, Ikuyuki and {Mizumoto}, Misaki and {Mizuno}, Tsunefumi and {Mori}, Koji and {Mukai}, Koji and {Murakami}, Hiroshi and {Mushotzky}, Richard and {Nakajima}, Hiroshi and {Nakazawa}, Kazuhiro and {Ness}, Jan-Uwe and {Nobukawa}, Kumiko and {Nobukawa}, Masayoshi and {Noda}, Hirofumi and {Odaka}, Hirokazu and {Ogawa}, Shoji and {Ogorzalek}, Anna and {Okajima}, Takashi and {Ota}, Naomi and {Paltani}, Stephane and {Petre}, Robert and {Plucinsky}, Paul and {Porter}, Frederick S. and {Pottschmidt}, Katja and {Sato}, Kosuke and {Sato}, Toshiki and {Sawada}, Makoto and {Seta}, Hiromi and {Shidatsu}, Megumi and {Simionescu}, Aurora and {Smith}, Randall and {Suzuki}, Hiromasa and {Szymkowiak}, Andrew and {Takahashi}, Hiromitsu and {Takeo}, Mai and {Tamagawa}, Toru and {Tamura}, Keisuke and {Tanaka}, Takaaki and {Tanimoto}, Atsushi and {Tashiro}, Makoto and {Terada}, Yukikatsu and {Terashima}, Yuichi and {Tsuboi}, Yohko and {Tsujimoto}, Masahiro and {Tsunemi}, Hiroshi and {Tsuru}, Takeshi and {Uchida}, Hiroyuki and {Uchida}, Nagomi and {Uchida}, Yuusuke and {Uchiyama}, Hideki and {Ueda}, Yoshihiro and {Uno}, Shinichiro and {Vink}, Jacco and {Watanabe}, Shin and {Williams}, Brian J. and {Yamada}, Satoshi and {Yamada}, Shinya and {Yamaguchi}, Hiroya and {Yamaoka}, Kazutaka and {Yamasaki}, Noriko and {Yamauchi}, Makoto and {Yamauchi}, Shigeo and {Yaqoob}, Tahir and {Yoneyama}, Tomokage and {Yoshida}, Tessei and {Yukita}, Mihoko and {Zhuravleva}, Irina and {Xiang}, Xin and {Minezaki}, Takeo and {Buhariwalla}, Margaret and {Gerolymatou}, Dimitra and {Hagen}, Scott},
        title = "{XRISM Spectroscopy of the Fe K{\ensuremath{\alpha}} Emission Line in the Seyfert Active Galactic Nucleus NGC 4151 Reveals the Disk, Broad-line Region, and Torus}",
      journal = {\apjl},
         year = 2024,
        month = sep,
       volume = {973},
       number = {1},
          eid = {L25},
        pages = {L25},
          doi = {10.3847/2041-8213/ad7397},
archivePrefix = {arXiv},
       eprint = {2408.14300},
 primaryClass = {astro-ph.HE},
       adsurl = {https://ui.adsabs.harvard.edu/abs/2024ApJ...973L..25X}
}

@ARTICLE{Miller2025ApJ...994L..10M,
       author = {{Miller}, Jon M. and {Xiang}, Xin and {Byun}, Doyee and {Behar}, Ehud and {Brenneman}, Laura and {Cackett}, Edward and {Costantini}, Elisa and {Gallo}, Luigi and {Horne}, Keith and {Kammoun}, Elias and {Li}, Chen and {Zoghbi}, Abderahmen},
        title = "{XRISM/Resolve Spectroscopy of the Central Engine in the Seyfert-1 AGN Mrk 279}",
      journal = {\apjl},
         year = 2025,
        month = nov,
       volume = {994},
       number = {1},
          eid = {L10},
        pages = {L10},
          doi = {10.3847/2041-8213/ae1606},
archivePrefix = {arXiv},
       eprint = {2510.20083},
 primaryClass = {astro-ph.HE},
       adsurl = {https://ui.adsabs.harvard.edu/abs/2025ApJ...994L..10M}
}

@ARTICLE{Noda2025PASJ...77S..10N,
       author = {{Noda}, Hirofumi and {Mori}, Koji and {Tomida}, Hiroshi and {Nakajima}, Hiroshi and {Tanaka}, Takaaki and {Murakami}, Hiroshi and {Uchida}, Hiroyuki and {Suzuki}, Hiromasa and {Kobayashi}, Shogo Benjamin and {Yoneyama}, Tomokage and {Hagino}, Kouichi and {Nobukawa}, Kumiko and {Uchiyama}, Hideki and {Nobukawa}, Masayoshi and {Matsumoto}, Hironori and {Tsuru}, Takeshi Go and {Yamauchi}, Makoto and {Hatsukade}, Isamu and {Odaka}, Hirokazu and {Kohmura}, Takayoshi and {Yamaoka}, Kazutaka and {Yoshida}, Tessei and {Kanemaru}, Yoshiaki and {Hiraga}, Junko and {Dotani}, Tadayasu and {Ozaki}, Masanobu and {Tsunemi}, Hiroshi and {Sato}, Jin and {Takaki}, Toshiyuki and {Terada}, Yuta and {Miyazaki}, Keitaro and {Kusunoki}, Kohei and {Otsuka}, Yoshinori and {Yokosu}, Haruhiko and {Yonemaru}, Wakana and {Ichikawa}, Kazuhiro and {Nakano}, Hanako and {Takemoto}, Reo and {Matsushima}, Tsukasa and {Urase}, Reika and {Kurashima}, Jun and {Fuchi}, Kotomi and {Hayakawa}, Kaito and {Fukuda}, Masahiro and {Kamei}, Takamitsu and {Asahina}, Yoh and {Inoue}, Shun and {Amano}, Yuki and {Aoki}, Yuma and {Ito}, Yamato and {Kamatani}, Tomoya and {Takayama}, Kouta and {Sako}, Takashi and {Yoshimoto}, Marina and {Shima}, Kohei and {Higuchi}, Mayu and {Ninoyu}, Kaito and {Aoki}, Daiki and {Tsunomachi}, Shun and {Hayashida}, Kiyoshi},
        title = "{Soft X-ray Imager of the Xtend system on board XRISM}",
      journal = {\pasj},
         year = 2025,
        month = sep,
       volume = {77},
        pages = {S10-S22},
          doi = {10.1093/pasj/psaf011},
archivePrefix = {arXiv},
       eprint = {2502.08030},
 primaryClass = {astro-ph.IM},
       adsurl = {https://ui.adsabs.harvard.edu/abs/2025PASJ...77S..10N}
}

@ARTICLE{Haynes1998AJ....115...62H,
       author = {{Haynes}, Martha P. and {Hogg}, David E. and {Maddalena}, Ronald J. and {Roberts}, Morton S. and {van Zee}, Liese},
        title = "{Asymmetry in High-Precision Global H i Profiles of Isolated Spiral Galaxies}",
      journal = {\aj},
         year = 1998,
        month = jan,
       volume = {115},
       number = {1},
        pages = {62-79},
          doi = {10.1086/300166},
       adsurl = {https://ui.adsabs.harvard.edu/abs/1998AJ....115...62H}
}

@ARTICLE{Fragile2024arXiv240410052F,
       author = {{Fragile}, P. Chris and {Liska}, Matthew},
        title = "{Tilted Accretion Disks}",
      journal = {arXiv e-prints},
         year = 2024,
        month = apr,
          eid = {arXiv:2404.10052},
        pages = {arXiv:2404.10052},
          doi = {10.48550/arXiv.2404.10052},
archivePrefix = {arXiv},
       eprint = {2404.10052},
 primaryClass = {astro-ph.HE},
       adsurl = {https://ui.adsabs.harvard.edu/abs/2024arXiv240410052F}
}

@ARTICLE{Lense1918PhyZ...19..156L,
       author = {{Lense}, Josef and {Thirring}, Hans},
        title = "{{\"U}ber den Einflu{\ss} der Eigenrotation der Zentralk{\"o}rper auf die Bewegung der Planeten und Monde nach der Einsteinschen Gravitationstheorie}",
      journal = {Physikalische Zeitschrift},
         year = 1918,
        month = jan,
       volume = {19},
        pages = {156},
       adsurl = {https://ui.adsabs.harvard.edu/abs/1918PhyZ...19..156L}
}

@ARTICLE{Bardeen1975ApJ...195L..65B,
       author = {{Bardeen}, James M. and {Petterson}, Jacobus A.},
        title = "{The Lense-Thirring Effect and Accretion Disks around Kerr Black Holes}",
      journal = {\apjl},
         year = 1975,
        month = jan,
       volume = {195},
        pages = {L65},
          doi = {10.1086/181711},
       adsurl = {https://ui.adsabs.harvard.edu/abs/1975ApJ...195L..65B}
}

@ARTICLE{Brenneman2025ApJ...995..200B,
       author = {{Brenneman}, Laura W. and {Wilkins}, Daniel R. and {Ogorza{\l}ek}, Anna and {Rogantini}, Daniele and {Fabian}, Andrew C. and {Garc{\'\i}a}, Javier A. and {Jur{\'a}{\v{n}}ov{\'a}}, Anna and {Mizumoto}, Misaki and {Noda}, Hirofumi and {Behar}, Ehud and {Boissay-Malaquin}, Rozenn and {Guainazzi}, Matteo and {Okajima}, Takashi and {Hoffman}, Erika and {Keshet}, Noa and {Kaastra}, Jelle and {Kara}, Erin and {Yamauchi}, Makoto},
        title = "{A Sharper View of the X-Ray Spectrum of MCG─6-30-15 with XRISM, XMM-Newton, and NuSTAR}",
      journal = {\apj},
         year = 2025,
        month = dec,
       volume = {995},
       number = {2},
          eid = {200},
        pages = {200},
          doi = {10.3847/1538-4357/ae1225},
archivePrefix = {arXiv},
       eprint = {2510.08926},
 primaryClass = {astro-ph.HE},
       adsurl = {https://ui.adsabs.harvard.edu/abs/2025ApJ...995..200B}
}

@ARTICLE{Tanaka1995Natur.375..659T,
       author = {{Tanaka}, Y. and {Nandra}, K. and {Fabian}, A.~C. and {Inoue}, H. and {Otani}, C. and {Dotani}, T. and {Hayashida}, K. and {Iwasawa}, K. and {Kii}, T. and {Kunieda}, H. and {Makino}, F. and {Matsuoka}, M.},
        title = "{Gravitationally redshifted emission implying an accretion disk and massive black hole in the active galaxy MCG-6-30-15}",
      journal = {\nat},
         year = 1995,
        month = jun,
       volume = {375},
       number = {6533},
        pages = {659-661},
          doi = {10.1038/375659a0},
       adsurl = {https://ui.adsabs.harvard.edu/abs/1995Natur.375..659T}
}

@ARTICLE{Fragile2005ApJ...623..347F,
       author = {{Fragile}, P. Chris and {Anninos}, Peter},
        title = "{Hydrodynamic Simulations of Tilted Thick-Disk Accretion onto a Kerr Black Hole}",
      journal = {\apj},
         year = 2005,
        month = apr,
       volume = {623},
       number = {1},
        pages = {347-361},
          doi = {10.1086/428433},
archivePrefix = {arXiv},
       eprint = {astro-ph/0403356},
 primaryClass = {astro-ph},
       adsurl = {https://ui.adsabs.harvard.edu/abs/2005ApJ...623..347F}
}

@ARTICLE{Papaloizou1995MNRAS.274..987P,
       author = {{Papaloizou}, John C.~B. and {Terquem}, Caroline},
        title = "{On the dynamics of tilted discs around young stars}",
      journal = {\mnras},
         year = 1995,
        month = jun,
       volume = {274},
       number = {4},
        pages = {987-1001},
          doi = {10.1093/mnras/274.4.987},
       adsurl = {https://ui.adsabs.harvard.edu/abs/1995MNRAS.274..987P}
}

@ARTICLE{Tashiro2025PASJ...77S...1T,
       author = {{Tashiro}, Makoto and {Kelley}, Richard and {Watanabe}, Shin and {Maejima}, Hironori and {Reichenthal}, Lillian and {Toda}, Kenichi and {Hartz}, Leslie and {Santovincenzo}, Andrea and {Matsushita}, Kyoko and {Yamaguchi}, Hiroya and {Petre}, Robert and {Williams}, Brian and {Guainazzi}, Matteo and {Costantini}, Elisa and {Takei}, Yoh and {Ishisaki}, Yoshitaka and {Fujimoto}, Ryuichi and {Henegar-Leon}, Joy and {Sneiderman}, Gary and {Tomida}, Hiroshi and {Mori}, Koji and {Nakajima}, Hiroshi and {Terada}, Yukikatsu and {Holland}, Matthew and {Loewenstein}, Michael and {Miller}, Eric and {Sawada}, Makoto and {Kallman}, Timothy and {Kaastra}, Jelle and {Done}, Chris and {Enoto}, Teruaki and {Bamba}, Aya and {Corrales}, Lia and {Ueda}, Yoshihiro and {Kara}, Erin and {Zhuravleva}, Irina and {Fujita}, Yutaka and {Arai}, Yoshitaka and {Audard}, Marc and {Awaki}, Hisamitsu and {Ballhausen}, Ralf and {Baluta}, Chris and {Bando}, Nobutaka and {Behar}, Ehud and {Bialas}, Thomas and {Boissay-Malaquin}, Rozenn and {Brenneman}, Laura and {Brown}, Gregory V. and {Chiao}, Meng and {Cumbee}, Renata and {de Vries}, Cor and {den Herder}, Jan-Willem and {D{\'\i}az Trigo}, Mar{\'\i}a and {DiPirro}, Michael and {Dotani}, Tadayasu and {Carrero}, Jacobo Ebrero and {Ebisawa}, Ken and {Eckart}, Megan and {Eckert}, Dominique and {Eguchi}, Satoshi and {Ezoe}, Yuichiro and {Ferrigno}, Carlo and {Foster}, Adam and {Fukazawa}, Yasushi and {Fukushima}, Kotaro and {Furuzawa}, Akihiro and {Gallo}, Luigi C. and {Garcia Martinez}, Javier and {Gorter}, Nathalie and {Grim}, Martin and {Gu}, Liyi and {Hagino}, Kouichi and {Hamaguchi}, Kenji and {Hatsukade}, Isamu and {Hayashi}, Katsuhiro and {Hayashi}, Takayuki and {Hell}, Natalie and {Hodges-Kluck}, Edmund and {Horiuchi}, Takafumi and {Hornschemeier}, Ann and {Hoshino}, Akio and {Ichinohe}, Yuto and {Ikuta}, Chisato and {Iizuka}, Ryo and {Ishi}, Daiki and {Ishida}, Manabu and {Ishihama}, Naoki and {Ishikawa}, Kumi and {Ishimura}, Kosei and {Jaffe}, Tess and {Katsuda}, Satoru and {Kanemaru}, Yoshiaki and {Kenyon}, Steven and {Kilbourne}, Caroline and {Kimball}, Mark and {Kitamoto}, Shunji and {Kobayashi}, Shogo and {Kohmura}, Takayoshi and {Kubota}, Aya and {Leutenegger}, Maurice A. and {Maeda}, Yoshitomo and {Markevitch}, Maxim and {Matsumoto}, Hironori and {Matsuzaki}, Keiichi and {McCammon}, Dan and {McLaughlin}, Brian and {McNamara}, Brian and {Mernier}, Fran{\c{c}}ois and {Miko}, Joseph and {Miller}, Jon M. and {Minesugi}, Kenji and {Mitani}, Shinji and {Mitsuishi}, Ikuyuki and {Mizumoto}, Misaki and {Mizuno}, Tsunefumi and {Mukai}, Koji and {Murakami}, Hiroshi and {Mushotzky}, Richard and {Nakazawa}, Kazuhiro and {Natsukari}, Chikara and {Ness}, Jan-Uwe and {Nigo}, Kenichiro and {Nishiyama}, Mari and {Nobukawa}, Kumiko and {Nobukawa}, Masayoshi and {Noda}, Hirofumi and {Odaka}, Hirokazu and {Ogawa}, Mina and {Ogawa}, Shoji and {Ogorzalek}, Anna and {Okajima}, Takashi and {Okamoto}, Atsushi and {Ota}, Naomi and {Ozaki}, Masanobu and {Paltani}, Stephane and {Plucinsky}, Paul and {Porter}, F. Scott and {Pottschmidt}, Katja and {Quero}, Jose Antonio and {Sasaki}, Takahiro and {Sato}, Kosuke and {Sato}, Rie and {Sato}, Toshiki and {Sato}, Yoichi and {Seta}, Hiromi and {Shida}, Maki and {Shidatsu}, Megumi and {Shigeto}, Shuhei and {Shipman}, Russel and {Shinozaki}, Keisuke and {Shirron}, Peter and {Simionescu}, Aurora and {Smith}, Randall K. and {Soong}, Yang and {Suzuki}, Hiromasa and {Szymkowiak}, Andrew and {Takahashi}, Hiromitsu and {Takeo}, Mai and {Tamagawa}, Toru and {Tamura}, Keisuke and {Tanaka}, Takaaki and {Tanimoto}, Atsushi and {Terashima}, Yuichi and {Tsuboi}, Yohko and {Tsujimoto}, Masahiro and {Tsunemi}, Hiroshi and {Tsuru}, Takeshi Go and {Uchida}, Hiroyuki and {Uchida}, Nagomi and {Uchida}, Yuusuke and {Uchiyama}, Hideki and {Uno}, Shinichiro and {Vink}, Jacco and {Witthoeft}, Michael and {Wolfs}, Rob and {Yamada}, Satoshi and {Yamada}, Shinya and {Yamaoka}, Kazutaka and {Yamasaki}, Noriko and {Yamauchi}, Makoto and {Yamauchi}, Shigeo and {Yanagase}, Keiichi and {Yaqoob}, Tahir and {Yasuda}, Susumu and {Yoneyama}, Tomokage and {Yoshida}, Tessei and {Yukita}, Mihoko},
        title = "{X-Ray Imaging and Spectroscopy Mission}",
      journal = {\pasj},
         year = 2025,
        month = sep,
       volume = {77},
        pages = {S1-S9},
          doi = {10.1093/pasj/psaf023},
       adsurl = {https://ui.adsabs.harvard.edu/abs/2025PASJ...77S...1T}
}

@ARTICLE{Ishisaki2025JATIS..11d2023I,
       author = {{Ishisaki}, Yoshitaka and {Kelley}, Richard L. and {Awaki}, Hisamitsu and {Balleza}, Jesus C. and {Barnstable}, Kim R. and {Bialas}, Thomas G. and {Boissay-Malaquin}, Rozenn and {Brown}, Gregory V. and {Canavan}, Edgar R. and {Cumbee}, Renata S. and {Carnahan}, Timothy M. and {Chiao}, Meng P. and {Comber}, Brian J. and {Costantini}, Elisa and {den Herder}, Jan-Willem and {Dercksen}, Johannes and {de Vries}, Cor P. and {DiPirro}, Michael J. and {Eckart}, Megan E. and {Ezoe}, Yuichiro and {Ferrigno}, Carlo and {Fujimoto}, Ryuichi and {Gorter}, Nathalie and {Graham}, Steven M. and {Grim}, Martin and {Hartz}, Leslie S. and {Hayakawa}, Ryota and {Hayashi}, Takayuki and {Hell}, Natalie and {Hoshino}, Akio and {Ichinohe}, Yuto and {Ishida}, Manabu and {Ishikawa}, Kumi and {James}, Bryan L. and {Kenyon}, Steven J. and {Kilbourne}, Caroline A. and {Kimball}, Mark O. and {Kitamoto}, Shunji and {Leutenegger}, Maurice A. and {Maeda}, Yoshitomo and {McCammon}, Dan and {Miko}, Joseph J. and {Mizumoto}, Misaki and {Noda}, Hirofumi and {Okajima}, Takashi and {Okamoto}, Atsushi and {Paltani}, Stephane and {Porter}, Frederick S. and {Sato}, Kosuke and {Sato}, Toshiki and {Sawada}, Makoto and {Shinozaki}, Keisuke and {Shipman}, Russell and {Shirron}, Peter J. and {Sneiderman}, Gary A. and {Soong}, Yang and {Szymkiewicz}, Richard and {Szymkowiak}, Andrew E. and {Takei}, Yoh and {Tamura}, Keisuke and {Tsujimoto}, Masahiro and {Uchida}, Yuusuke and {Wasserzug}, Stephen and {Witthoeft}, Michael C. and {Wolfs}, Rob and {Yamada}, Shinya and {Yasuda}, Susumu},
        title = "{Resolve instrument onboard XRISM: design, integration, and instrument test results}",
      journal = {Journal of Astronomical Telescopes, Instruments, and Systems},
         year = 2025,
        month = oct,
       volume = {11},
          eid = {042023},
        pages = {042023},
          doi = {10.1117/1.JATIS.11.4.042023},
       adsurl = {https://ui.adsabs.harvard.edu/abs/2025JATIS..11d2023I}
}

@ARTICLE{Kelley2025JATIS..11d2026K,
       author = {{Kelley}, Richard L. and {Ishisaki}, Yoshitaka and {Costantini}, Elisa and {Awaki}, Hisamitsu and {Balleza}, Jesus C. and {Barnstable}, Kim R. and {Bialas}, Thomas G. and {Boissay-Malaquin}, Rozenn and {Brown}, Gregory V. and {Canavan}, Edgar R. and {Carnahan}, Timothy M. and {Chiao}, Meng P. and {Comber}, Brian J. and {Cumbee}, Renata S. and {den Herder}, Jan-Willem and {Dercksen}, Johannes and {de Vries}, Cor P. and {DiPirro}, Michael J. and {Eckart}, Megan E. and {Ezoe}, Yuichiro and {Ferrigno}, Carlo and {Fujimoto}, Ryuichi and {Gorter}, Nathalie and {Graham}, Steven M. and {Grim}, Martin and {Hartz}, Leslie S. and {Hayakawa}, Ryota and {Hayashi}, Takayuki and {Hell}, Natalie and {Ichinohe}, Yuto and {Ishi}, Daiki and {Ishida}, Manabu and {Ishikawa}, Kumi and {James}, Bryan L. and {Kanemaru}, Yoshiaki and {Kenyon}, Steven J. and {Kilbourne}, Caroline A. and {Kimball}, Mark O. and {Kitamoto}, Shunji and {Leutenegger}, Maurice A. and {Maeda}, Yoshitomo and {McCammon}, Dan and {McLaughlin}, Brian J. and {Miko}, Joseph J. and {van der Meer}, Erik and {Mizumoto}, Misaki and {Noda}, Hirofumi and {Okajima}, Takashi and {Okamoto}, Atsushi and {Paltani}, Stephane and {Porter}, Frederick S. and {Reichenthal}, Lillian S. and {Sato}, Kosuke and {Sato}, Toshiki and {Sato}, Yoichi and {Sawada}, Makoto and {Shinozaki}, Keisuke and {Shipman}, Russell and {Shirron}, Peter J. and {Sneiderman}, Gary A. and {Soong}, Yang and {Szymkiewicz}, Richard and {Szymkowiak}, Andrew E. and {Takei}, Yoh and {Takeo}, Mai and {Tamura}, Keisuke and {Tsujimoto}, Masahiro and {Uchida}, Yuusuke and {Wasserzug}, Stephen and {Witthoeft}, Michael C. and {Wolfs}, Rob and {Yamada}, Shinya and {Yamasaki}, Noriko Y. and {Yasuda}, Susumu},
        title = "{Resolve instrument onboard the X-Ray Imaging and Spectroscopy Mission}",
      journal = {Journal of Astronomical Telescopes, Instruments, and Systems},
         year = 2025,
        month = oct,
       volume = {11},
          eid = {042026},
        pages = {042026},
          doi = {10.1117/1.JATIS.11.4.042026},
       adsurl = {https://ui.adsabs.harvard.edu/abs/2025JATIS..11d2026K}
}

@INPROCEEDINGS{2024SPIE13093E..1PE,
       author = {{Eckart}, Megan E. and {Brown}, Gregory V. and {Chiao}, Meng P. and {Cumbee}, Renata S. and {Fujimoto}, Ryuichi and {Hell}, Natalie and {Hoshino}, Akio and {Ishisaki}, Yoshitaka and {Kelley}, Richard L. and {Kenyon}, Steven J. and {Kilbourne}, Caroline A. and {Kitamoto}, Shunji and {Leutenegger}, Maurice A. and {Lockard}, Tom and {Loewenstein}, Michael and {Magee}, Edward W. and {Mizumoto}, Misaki and {Porter}, F. Scott and {Sato}, Kosuke and {Sawada}, Makoto and {Shah}, Chintan and {Shipman}, Russell F. and {Sneiderman}, Gary A. and {Takei}, Yoh and {Tsujimoto}, Masahiro and {de Vries}, Cor P. and {Watanabe}, Tomomi and {Witthoeft}, Michael C. and {Wolfs}, Rob and {Yamada}, Shinya and {Yaqoob}, Tahir},
        title = "{Energy gain scale calibration of the XRISM Resolve microcalorimeter spectrometer: ground calibration results and on-orbit comparison}",
    booktitle = {Space Telescopes and Instrumentation 2024: Ultraviolet to Gamma Ray},
         year = 2024,
       editor = {{den Herder}, Jan-Willem A. and {Nikzad}, Shouleh and {Nakazawa}, Kazuhiro},
       series = {Society of Photo-Optical Instrumentation Engineers (SPIE) Conference Series},
       volume = {13093},
        month = aug,
          eid = {130931P},
        pages = {130931P},
          doi = {10.1117/12.3019276},
       adsurl = {https://ui.adsabs.harvard.edu/abs/2024SPIE13093E..1PE}
}

@ARTICLE{Brenneman2006ApJ...652.1028B,
       author = {{Brenneman}, Laura W. and {Reynolds}, Christopher S.},
        title = "{Constraining Black Hole Spin via X-Ray Spectroscopy}",
      journal = {\apj},
         year = 2006,
        month = dec,
       volume = {652},
       number = {2},
        pages = {1028-1043},
          doi = {10.1086/508146},
archivePrefix = {arXiv},
       eprint = {astro-ph/0608502},
 primaryClass = {astro-ph},
       adsurl = {https://ui.adsabs.harvard.edu/abs/2006ApJ...652.1028B}
}

@ARTICLE{Fabian1989MNRAS.238..729F,
       author = {{Fabian}, A.~C. and {Rees}, M.~J. and {Stella}, L. and {White}, N.~E.},
        title = "{X-ray fluorescence from the inner disc in Cygnus X-1.}",
      journal = {\mnras},
         year = 1989,
        month = may,
       volume = {238},
        pages = {729-736},
          doi = {10.1093/mnras/238.3.729},
       adsurl = {https://ui.adsabs.harvard.edu/abs/1989MNRAS.238..729F}
}

@ARTICLE{Moran2005AJ....129.2108M,
       author = {{Moran}, Edward C. and {Eracleous}, Michael and {Leighly}, Karen M. and {Chartas}, George and {Filippenko}, Alexei V. and {Ho}, Luis C. and {Blanco}, Philip R.},
        title = "{Extreme X-Ray Behavior of the Low-Luminosity Active Nucleus in NGC 4395}",
      journal = {\aj},
         year = 2005,
        month = may,
       volume = {129},
       number = {5},
        pages = {2108-2118},
          doi = {10.1086/429522},
archivePrefix = {arXiv},
       eprint = {astro-ph/0502109},
 primaryClass = {astro-ph},
       adsurl = {https://ui.adsabs.harvard.edu/abs/2005AJ....129.2108M}
}

@ARTICLE{Kaastra2016A&A...587A.151K,
       author = {{Kaastra}, J.~S. and {Bleeker}, J.~A.~M.},
        title = "{Optimal binning of X-ray spectra and response matrix design}",
      journal = {\aap},
         year = 2016,
        month = mar,
       volume = {587},
          eid = {A151},
        pages = {A151},
          doi = {10.1051/0004-6361/201527395},
archivePrefix = {arXiv},
       eprint = {1601.05309},
 primaryClass = {astro-ph.IM},
       adsurl = {https://ui.adsabs.harvard.edu/abs/2016A&A...587A.151K}
}

@ARTICLE{HI4PI2016A&A...594A.116H,
       author = {{HI4PI Collaboration} and {Ben Bekhti}, N. and {Fl{\"o}er}, L. and {Keller}, R. and {Kerp}, J. and {Lenz}, D. and {Winkel}, B. and {Bailin}, J. and {Calabretta}, M.~R. and {Dedes}, L. and {Ford}, H.~A. and {Gibson}, B.~K. and {Haud}, U. and {Janowiecki}, S. and {Kalberla}, P.~M.~W. and {Lockman}, F.~J. and {McClure-Griffiths}, N.~M. and {Murphy}, T. and {Nakanishi}, H. and {Pisano}, D.~J. and {Staveley-Smith}, L.},
        title = "{HI4PI: A full-sky H I survey based on EBHIS and GASS}",
      journal = {\aap},
         year = 2016,
        month = oct,
       volume = {594},
          eid = {A116},
        pages = {A116},
          doi = {10.1051/0004-6361/201629178},
archivePrefix = {arXiv},
       eprint = {1610.06175},
 primaryClass = {astro-ph.GA},
       adsurl = {https://ui.adsabs.harvard.edu/abs/2016A&A...594A.116H}
}

@ARTICLE{Nardini2011MNRAS.417.2571N,
       author = {{Nardini}, E. and {Risaliti}, G.},
        title = "{The effects of X-ray absorption variability in NGC 4395}",
      journal = {\mnras},
         year = 2011,
        month = nov,
       volume = {417},
       number = {4},
        pages = {2571-2576},
          doi = {10.1111/j.1365-2966.2011.19423.x},
archivePrefix = {arXiv},
       eprint = {1107.2405},
 primaryClass = {astro-ph.CO},
       adsurl = {https://ui.adsabs.harvard.edu/abs/2011MNRAS.417.2571N}
}

@ARTICLE{Kammoun2019ApJ...886..145K,
       author = {{Kammoun}, E.~S. and {Nardini}, E. and {Zoghbi}, A. and {Miller}, J.~M. and {Cackett}, E.~M. and {Gallo}, E. and {Reynolds}, M.~T. and {Risaliti}, G. and {Barret}, D. and {Brandt}, W.~N. and {Brenneman}, L.~W. and {Kaastra}, J.~S. and {Koss}, M. and {Lohfink}, A.~M. and {Mushotzky}, R.~F. and {Raymond}, J. and {Stern}, D.},
        title = "{The Nature of the Broadband X-Ray Variability in the Dwarf Seyfert Galaxy NGC 4395}",
      journal = {\apj},
         year = 2019,
        month = dec,
       volume = {886},
       number = {2},
          eid = {145},
        pages = {145},
          doi = {10.3847/1538-4357/ab5110},
archivePrefix = {arXiv},
       eprint = {1910.11317},
 primaryClass = {astro-ph.HE},
       adsurl = {https://ui.adsabs.harvard.edu/abs/2019ApJ...886..145K}
}

@ARTICLE{denBrok2015ApJ...809..101D,
       author = {{den Brok}, Mark and {Seth}, Anil C. and {Barth}, Aaron J. and {Carson}, Daniel J. and {Neumayer}, Nadine and {Cappellari}, Michele and {Debattista}, Victor P. and {Ho}, Luis C. and {Hood}, Carol E. and {McDermid}, Richard M.},
        title = "{Measuring the Mass of the Central Black Hole in the Bulgeless Galaxy NGC 4395 from Gas Dynamical Modeling}",
      journal = {\apj},
         year = 2015,
        month = aug,
       volume = {809},
       number = {1},
          eid = {101},
        pages = {101},
          doi = {10.1088/0004-637X/809/1/101},
archivePrefix = {arXiv},
       eprint = {1507.04358},
 primaryClass = {astro-ph.GA},
       adsurl = {https://ui.adsabs.harvard.edu/abs/2015ApJ...809..101D}
}

@inbook{Burnham_2002, address={New York, NY}, title={Information and Likelihood Theory: A Basis for Model Selection and Inference}, ISBN={978-0-387-22456-5}, url={https://doi.org/10.1007/978-0-387-22456-5_2}, DOI={10.1007/978-0-387-22456-5_2}, abstractNote={Full reality cannot be included in a model; thus we seek a good model to approximate the effects or factors supported by the empirical data. The selection of an appropriate approximating model is critical to statistical inference from many types of empirical data. This chapter introduces concepts from information theory (see Guiasu 1977), which has been a discipline only since the mid-1940s and covers a variety of theories and methods that are fundamental to many of the sciences (see Cover and Thomas 1991 for an exciting overview; Figure 2.1 is produced from their book and shows their view of the relationship of information theory to several other fields). In particular, the Kullback—Leibler “distance,” or “information,” between two models (Kull-back and Leibler 1951) is introduced, discussed, and linked to Boltzmann’s entropy in this chapter. Akaike (1973) found a simple relationship between the Kullback—Leibler distance and Fisher’s maximized log-likelihood function (see deLeeuw 1992 for a brief review). This relationship leads to a simple, effective, and very general methodology for selecting a parsimonious model for the analysis of empirical data.}, booktitle={Model Selection and Multimodel Inference: A Practical Information-Theoretic Approach}, publisher={Springer}, author={Burnham, Kenneth P. and Anderson, David R.}, year={2002}, pages={49–97}, language={en} }

@ARTICLE{Akaike1974ITAC...19..716A,
       author = {{Akaike}, H.},
        title = "{A New Look at the Statistical Model Identification}",
      journal = {IEEE Transactions on Automatic Control},
         year = 1974,
        month = jan,
       volume = {19},
        pages = {716-723},
          doi = {10.1109/TAC.1974.1100705},
       adsurl = {https://ui.adsabs.harvard.edu/abs/1974ITAC...19..716A}
}

@ARTICLE{Fujiwara2026arXiv260406719F,
       author = {{Fujiwara}, Kanta and {Ueda}, Yoshihiro and {Ogawa}, Shoji and {Nakatani}, Yuya and {Miller}, Jon M. and {Okajima}, Takashi and {Kawamuro}, Taiki and {Boorman}, Peter G. and {Gallo}, Luigi and {Mizumoto}, Misaki and {Mushotzky}, Richard and {Noda}, Hirofumi and {Terashima}, Yuichi and {Tombesi}, Francesco and {Vander Meulen}, Bert and {Yamada}, Satoshi},
        title = "{Complex Nuclear Structure in Seyfert 2 Galaxy NGC 4388 Revealed by XRISM Observation}",
      journal = {arXiv e-prints},
         year = 2026,
        month = apr,
          eid = {arXiv:2604.06719},
        pages = {arXiv:2604.06719},
          doi = {10.48550/arXiv.2604.06719},
archivePrefix = {arXiv},
       eprint = {2604.06719},
 primaryClass = {astro-ph.HE},
       adsurl = {https://ui.adsabs.harvard.edu/abs/2026arXiv260406719F}
}

@ARTICLE{Wilkins2026arXiv260409761W,
       author = {{Wilkins}, D.~R. and {Brenneman}, L.~W. and {Ogorzalek}, A. and {Fabian}, A.~C. and {Behar}, E. and {Boissay-Malaquin}, R. and {Garcia}, J.~A. and {Hoffman}, E.~B. and {Juranova}, A. and {Rogantini}, D.},
        title = "{Time-resolved XRISM spectroscopy reveals the evolution and structure of the corona in MCG-6-30-15}",
      journal = {arXiv e-prints},
         year = 2026,
        month = apr,
          eid = {arXiv:2604.09761},
        pages = {arXiv:2604.09761},
          doi = {10.48550/arXiv.2604.09761},
archivePrefix = {arXiv},
       eprint = {2604.09761},
 primaryClass = {astro-ph.HE},
       adsurl = {https://ui.adsabs.harvard.edu/abs/2026arXiv260409761W}
}

@ARTICLE{XRISM2026NatAs.tmp...63X,
       author = {{XRISM Collaboration} and {Audard}, Marc and {Awaki}, Hisamitsu and {Ballhausen}, Ralf and {Bamba}, Aya and {Behar}, Ehud and {Boissay-Malaquin}, Rozenn and {Brenneman}, Laura and {Brown}, Gregory V. and {Corrales}, Lia and {Costantini}, Elisa and {Cumbee}, Renata and {D{\'\i}az Trigo}, Mar{\'\i}a and {Done}, Chris and {Dotani}, Tadayasu and {Ebisawa}, Ken and {Eckart}, Megan E. and {Eckert}, Dominique and {Enoto}, Teruaki and {Eguchi}, Satoshi and {Ezoe}, Yuichiro and {Foster}, Adam and {Fujimoto}, Ryuichi and {Fujita}, Yutaka and {Fukazawa}, Yasushi and {Fukushima}, Kotaro and {Furuzawa}, Akihiro and {Gallo}, Luigi and {Garcia}, Javier A. and {Gu}, Liyi and {Guainazzi}, Matteo and {Hagino}, Kouichi and {Hamaguchi}, Kenji and {Hatsukade}, Isamu and {Hayashi}, Katsuhiro and {Hayashi}, Takayuki and {Hell}, Natalie and {Hodges-Kluck}, Edmund and {Hornschemeier}, Ann and {Ichinohe}, Yuto and {Ishi}, Daiki and {Ishida}, Manabu and {Ishikawa}, Kumi and {Ishisaki}, Yoshitaka and {Kaastra}, Jelle and {Kallman}, Timothy and {Kara}, Erin and {Katsuda}, Satoru and {Kanemaru}, Yoshiaki and {Kelley}, Richard and {Kilbourne}, Caroline and {Kitamoto}, Shunji and {Kobayashi}, Shogo and {Kohmura}, Takayoshi and {Kubota}, Aya and {Leutenegger}, Maurice and {Loewenstein}, Michael and {Maeda}, Yoshitomo and {Markevitch}, Maxim and {Matsumoto}, Hironori and {Matsushita}, Kyoko and {McCammon}, Dan and {McNamara}, Brian and {Mernier}, Fran{\c{c}}ois and {Miller}, Eric D. and {Miller}, Jon M. and {Mitsuishi}, Ikuyuki and {Mizumoto}, Misaki and {Mizuno}, Tsunefumi and {Mori}, Koji and {Mukai}, Koji and {Murakami}, Hiroshi and {Mushotzky}, Richard and {Nakajima}, Hiroshi and {Nakazawa}, Kazuhiro and {Ness}, Jan-Uwe and {Nobukawa}, Kumiko and {Nobukawa}, Masayoshi and {Noda}, Hirofumi and {Odaka}, Hirokazu and {Ogawa}, Shoji and {Ogorzalek}, Anna and {Okajima}, Takashi and {Ota}, Naomi and {Paltani}, Stephane and {Petre}, Robert and {Plucinsky}, Paul and {Porter}, Frederick Scott and {Pottschmidt}, Katja and {Sato}, Kosuke and {Sato}, Toshiki and {Sawada}, Makoto and {Seta}, Hiromi and {Shidatsu}, Megumi and {Simionescu}, Aurora and {Smith}, Randall and {Suzuki}, Hiromasa and {Szymkowiak}, Andrew and {Takahashi}, Hiromitsu and {Takeo}, Mai and {Tamagawa}, Toru and {Tamura}, Keisuke and {Tanaka}, Takaaki and {Tanimoto}, Atsushi and {Tashiro}, Makoto and {Terada}, Yukikatsu and {Terashima}, Yuichi and {Tsuboi}, Yohko and {Tsujimoto}, Masahiro and {Tsunemi}, Hiroshi and {Tsuru}, Takeshi G. and {T{\"u}mer}, Ay{\textcommabelow s}eg{\"u}l and {Uchida}, Hiroyuki and {Uchida}, Nagomi and {Uchida}, Yuusuke and {Uchiyama}, Hideki and {Ueda}, Yoshihiro and {Uno}, Shinichiro and {Vink}, Jacco and {Watanabe}, Shin and {Williams}, Brian J. and {Yamada}, Satoshi and {Yamada}, Shinya and {Yamaguchi}, Hiroya and {Yamaoka}, Kazutaka and {Yamasaki}, Noriko and {Yamauchi}, Makoto and {Yamauchi}, Shigeo and {Yaqoob}, Tahir and {Yoneyama}, Tomokage and {Yoshida}, Tessei and {Yukita}, Mihoko and {Zhuravleva}, Irina and {Fujiwara}, Kanta and {Izumi}, Takuma and {Kawamuro}, Taiki and {Maeda}, Keiichi and {Nakatani}, Yuya and {Paerels}, Frits and {Uematsu}, Ryosuke and {Vander Meulen}, Bert},
        title = "{Accurate determination of chemical abundances near a supermassive black hole}",
      journal = {Nature Astronomy},
         year = 2026,
        month = mar,
          doi = {10.1038/s41550-026-02817-6},
archivePrefix = {arXiv},
       eprint = {2603.29748},
 primaryClass = {astro-ph.HE},
       adsurl = {https://ui.adsabs.harvard.edu/abs/2026NatAs.tmp...63X}
}
\bibliographystyle{aasjournalv7}



\end{document}